\documentclass[a4paper,fleqn,usenatbib]{mnras}
\usepackage{newtxtext,newtxmath}

\usepackage[T1]{fontenc}
\usepackage{ae,aecompl}

\usepackage{graphicx}	
\usepackage{amsmath}	
\usepackage{amssymb}	
\usepackage{times}
\usepackage{gensymb}
\usepackage{pifont}
\usepackage{placeins}
\usepackage{textcomp}
\newcommand{\cmark}{\ding{51}}
\title[The kinematics of OB associations]{Not all stars form in clusters -- measuring the kinematics of OB associations with {\it Gaia}}

\author[J. L. Ward \& J. M. D. Kruijssen]{
Jacob L. Ward\thanks{E-mail: ward@uni-heidelberg.de (JLW)}
and J. M. Diederik Kruijssen
\\
Astronomisches Rechen-Institut, Zentrum f\"{u}r Astronomie der Universit\"{a}t Heidelberg, M\"{o}nchhofstra{\ss}e 12-14, D-69120 Heidelberg, Germany
}

\date{Accepted 2018 January 10. Received 2017 November 27; in original form 2017 September 22.}

\pubyear{2017}

\begin{document}
\label{firstpage}
\pagerange{\pageref{firstpage}--\pageref{lastpage}}
\maketitle

\begin{abstract}
It is often stated that star clusters are the fundamental units of star formation and that most (if not all) stars form in dense stellar clusters. In this monolithic formation scenario, low density OB associations are formed from the expansion of gravitationally bound clusters following gas expulsion due to stellar feedback. 
 $N$-body simulations of this process show that OB associations formed this way retain signs of expansion and elevated radial anisotropy over tens of Myr.
However, recent theoretical and observational studies suggest that star formation is a hierarchical process, following the fractal nature of natal molecular clouds and allowing the formation of large-scale associations in-situ. We distinguish between these two scenarios by characterising the kinematics of OB associations using the Tycho-{\it Gaia} Astrometric Solution catalogue. To this end, we quantify four key kinematic diagnostics: the number ratio of stars with positive radial velocities to those with negative radial velocities, the median radial velocity,  the median radial velocity normalised by the tangential velocity, and the radial anisotropy parameter. Each quantity presents a useful diagnostic of whether the association was more compact in the past. We compare these diagnostics to models representing random motion and the expanding products of monolithic cluster formation. None of these diagnostics show evidence of expansion, either from a single cluster or multiple clusters, and the observed kinematics are better represented by a random velocity distribution. This result favours the hierarchical star formation model in which a minority of stars forms in bound clusters and large-scale, hierarchically-structured associations are formed in-situ.
\end{abstract}




\section{Introduction}


It is often assumed that most (if not all) stars are formed in dense, embedded, gravitationally bound clusters.
In this monolithic model of star formation, observed low density associations of OB stars must have been formed as single dense stellar clusters (i.e. \textquotedblleft singularly monolithic\textquotedblright) or multiple stellar clusters (i.e. \textquotedblleft multiply monolithic\textquotedblright) that subsequently underwent a period of expansion to form the large scale structures visible today \citep{LadaLada1991,Brown1997,Kroupa2001}. It is postulated that following the formation of stars, gas is expelled from clusters through feedback, rendering the clusters super-virial and allowing dispersal \citep{Hills1980,Goodwin2006,Baumgardt2007}. 
This mechanism has often been used to explain the low proportion of gas-free, gravitationally bound clusters clusters after a few Myr.
However, star formation is observed to take place in a wide range of environments including large scale hierarchical structures and isolated young stellar objects which are not associated with clusters (e.g. \citealt{Gomez1993, Lamb2010,Allen2007,Gutermuth2008, Evans2009}).

In an alternative,  hierarchical model of star formation,  (e.g. \citealt{Elmegreen2002, Elmegreen2008, Bastian2007,Bonnell2011,Kruijssen2012b}) 
 stars form in a substructured and scale-free distribution of initial conditions (as expected from a supersonically turbulent interstellar medium, ISM), which can naturally explain the wide range of observed stellar densities. 
  \citet{Bressert2010} find that present-day, nearby young stellar objects (YSOs) are observed over a continuous density distribution, without features corresponding to individual clusters, concluding that this is likely a result of star formation occurring over a continuous density distribution. However, this conclusion has been contested by studies showing that expanding clusters can reproduce a similar result \citep{Gieles2012,Parker2012,Pfalzner2012}.
 Recent observational studies of the progenitors of young massive clusters (YMCs) have concluded that the density is insufficient to form the observed stellar densities of YMCs in-situ and that therefore the mass must become more centrally concentrated as the (proto)cluster evolves (\citealt{Walker2015,Walker2016}, see also the review of \citealt{Longmore2014}), in strong contrast with the idea of gas expulsion-driven expansion towards the present-day densities.
 
 Recent theoretical work has also called into question the effectiveness of gas expulsion as a means of cluster disruption, finding that the highest-density regions achieve gas exhaustion rather than gas expulsion, due to the short free-fall times and correspondingly high star formation efficiencies \citep{Kruijssen2012c,Girichidis2012}. As a result, the dynamical effect of gas expulsion is small. This work is supported by the observations presented by \citet{Ginsburg2016}, who find that the W51 protoclusters are evolving towards a state of gas exhaustion rather than gas expulsion.
The above concepts have been combined in the analytical theory for the fraction of star formation occurring in bound clusters (i.e. the cluster formation efficiency) by \citet{Kruijssen2012b}, who translates the local star formation efficiency to a bound fraction {by integrating over the continuous density spectrum of the ISM. The model predicts that the fraction of star formation occurring in bound clusters increases with gas pressure and surface density, leading to the conclusion that clusters are a possible outcome of star formation, rather than a fundamental unit.

The monolithic model of star formation firmly predicts that, if gas expulsion is a rapid process (operating on a time-scale smaller than one crossing time), the stars outside of the Lagrangian radius containing 20\% of the stars in a cluster will acquire strongly radially anisotropic velocities. For the first 20\,Myr, the $N$-body simulations of \citet{Baumgardt2007} predict that expanding clusters are strongly radially anisotropic and super-virial. This anisotropy initially takes values in the range $\beta=0.4$--$1$, but is expected to persist to a value of $\beta\sim0.2$ up to hundreds of initial crossing times (see Figures 8 and 9 in \citealt{Baumgardt2007}).
If clusters expand into transient associations before ultimately dispersing into the Galactic field, then we would expect OB associations to typically exhibit strongly radially anisotropic velocity fields with (on average) positive outward radial motions.
Regardless of the number of expanding clusters, and their geometry, expansion of clusters within an association must result in a net expansion of the association.
Therefore, observing the dynamics of present-day OB associations offers a powerful diagnostic to distinguish between the two paradigms of star and cluster formation, i.e. monolithic or hierarchical star formation.

This is not the first paper aiming to address this question by studying the dynamics of OB associations.
\citet{Preibisch1999} presented a detailed analysis of the distribution and kinematics in the Upper Sco OB association; however, they make no conclusions as to the origin of the association.
In a detailed analysis of the structure and kinematics of the OB association Cygnus OB2,  \citet{Wright2014,Wright2016} found that the level of dynamical evolution experienced by this association is low, concluding that Cygnus OB2 most likely formed as a highly sub-structured and globally unbound system, i.e.~it was always an association.
While previous studies have targeted individual associations, our study capitalises on the unprecedented surveying power of Gaia to greatly expand the sample of OB associations, providing a systematic and comprehensive analysis.

%

In this work we test the hypothesis that all OB associations result from the expansion of one or more gravitationally bound clusters. We focus only on the generalised behaviour of OB associations and do not fit models to any specific OB associations.  The main goal of this paper is to assess the validity of the singularly monolithic and multiply monolithic models of star formation as defined in this introduction.
We test the hypothesis that all OB associations are expanded (massive) star clusters using the first {\it Gaia} \citep{GAIA2016} data release (DR1, \citealt{GAIADR12016}).
Using the Tycho-{\it Gaia} Astrometric Solution (TGAS, \citealt{Michalik2015}) catalogue, we have carried out a 5-dimensional (position, distance, proper motion) association member selection of stars in the vicinities of known OB associations. We have performed a 4-dimensional analysis for each association, measuring the positions and velocities in the plane of the sky in order to determine whether the present day OB associations are indeed undergoing gas expulsion-driven expansion.
In Section 2, we introduce our sample selection and association membership criteria as well as our initial data reduction and the generation of model data sets with which to compare the observed data.
In Section 3, we quantify four key kinematic properties of the OB associations and compare these to model distributions.
 The implications of these results and remaining caveats are discussed in Section 4. Finally, our conclusions are summarised in Section 5.


\section{Sample selection, data reduction and model distributions}

Accurate and precise distances towards OB association members are crucial for both membership selection and the analysis of dynamical properties of those associations. This study is therefore limited to a strictly enforced 3$\sigma$ confidence limit on the distances (i.e.~$\sigma_D\leq D/3$) derived from the TGAS catalogue parallaxes. 
We obtained parallaxes and corresponding uncertainties from the {\it Gaia} archive\footnote{https://gea.esac.esa.int/archive/} for the entire TGAS catalogue (entries for $\sim$2 million stars). From these parallax measurements, we calculate distances and distance uncertainties. 

Figure \ref{tgas_distance_fig} shows the uncertainty in distance relative to the distance (derived from TGAS parallaxes) against distance in kpc for the entire TGAS catalogue. 
Beyond a distance of 1.6\,kpc, no stars have distances determined to the precision required in this study.
This motivates an absolute distance cut of 1.6\,kpc as no stars have distances measured to a 3$\sigma$ confidence interval beyond this distance. Only OB associations with literature distances of less than or equal to 1.6\,kpc are therefore considered in the remainder of this work.
For OB associations within this distance, we only consider the stars with distance uncertainties smaller than 1/3 of the distance.
This defines the parent sample from which OB associations will be extracted based on their clustering in 5-dimensional phase space, yielding much smaller numbers of stars ($N_*=100$--$400$). 

\begin{figure}
	\includegraphics[width=0.99\linewidth]{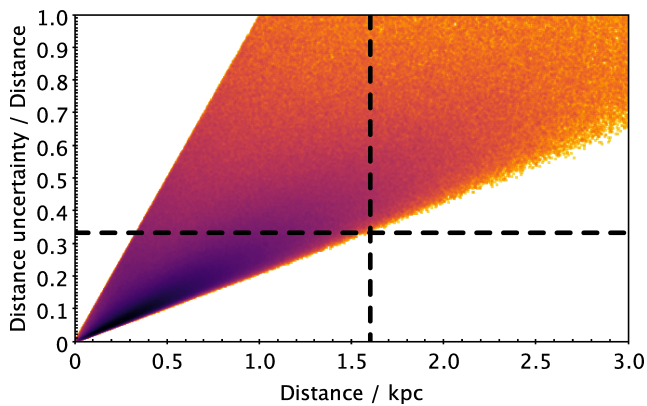}
	\caption{\label{tgas_distance_fig} Relative uncertainties in the derived distances against distance based on parallax measurements for the entire TGAS catalogue. The dashed lines mark the 3$\sigma$ distance uncertainty cut-off imposed in this work, motivating a maximum distance cut of 1.6\,kpc. Only those stars that fall in the lower left quadrant of this Figure are considered in this study. This defines the parent sample from which OB associations will be extracted based on their clustering in 5-dimensional phase space, yielding much smaller numbers of stars ($N_*=100$-$400$).}
\end{figure}

\subsection{TGAS data}

\label{TGAS_data}

We use data from the TGAS catalogue of stellar parallaxes and proper motions to characterise the kinematic properties of nearby OB associations.
We obtained parallaxes and proper motions (along with corresponding uncertainties) for potential members of each of the 43 OB associations within 1.6\,kpc of the sun based on the distances of \citet{Melnik2009}. These data were obtained using a cone search with a radius of 150\,pc at the literature distance for each association.
For each OB association, data for all stars with distances constrained to less than 3$\sigma$ have been excluded from the remainder of our study.
The remaining sample was then cross-referenced with the Simbad database, determining the nearest counterparts within a positional tolerance of 1$^{\prime\prime}$ to define our sample of OB-type stars. A cut has been imposed to exclude sources with distances greater than 150\,pc from the expected association distances from \citet{Melnik2009}. 

Using the distances determined from the TGAS parallaxes, we  calculate 2-dimensional physical distances ($X$,$Y$) in the plane of the sky from the median position of the OB stars in the sample in pc. Likewise, the same distances are used to convert observed proper motions from mas/yr to physical velocities in km\,s$^{-1}$. 
We define over-abundances of OB stars as regions with peak number densities of at least three times the observed background level of O- and B- type stars measured across the 150\,pc radius region with bin sizes of 2\,pc.
We then determine the position and size of the associations in each axis by fitting Gaussian profiles to the over-densities in each of the 5 dimensions available: $X$, $Y$, $D$, $v_x$, and $v_y$.
 
 Association members are selected as satisfying 
 $|x| < \sigma_X$ and $|y| < \sigma_Y$, where $x \equiv X-\mu_X$ and $y \equiv Y-\mu_Y$.
  A strict 1$\sigma$ criterion is imposed where possible in $X$ and $Y$ in order to minimise the likelihood of selecting contaminating field stars as association members. In cases where the 1$\sigma$ criterion did  not yield at least 100 stars including at least 10 O- or B- type stars, a 3$\sigma$ criterion was imposed and if insufficient numbers of stars were selected within 3$\sigma$ the association was excluded from the sample. 
   Similarly, we impose a line-of-sight distance selection criterion of $|Z| < (\Delta D^{2} + \sigma_Z^{2})^{1/2}$  where $Z \equiv D - \mu_{D} $ and $\Delta D$ is the uncertainty as derived from the uncertainty in the TGAS parallax measurement. This then removes the outliers in line-of-sight distance from the sample. The dispersions used to carry out the above selection criteria are listed in Table \ref{disp_tbl}.

   \begin{table}
   	\caption{\label{disp_tbl} Distances (D) and measured dispersions for each cluster in the five dimensions used in the selection criteria of Section 2.1: dispersions in the plane of the sky ($\sigma_{X}$,$\sigma_{Y}$), dispersion in the line-of-sight distance derived from TGAS parallaxes ($\sigma_{Z}$), and velocity dispersions in the plane of the sky ($\sigma_{v_{X}}$, $\sigma_{v_{Y}}$).}
   	\begin{center}
   		\begin{tabular}{l c c c c c c}
   			\hline
   		OB assoc. & D & $\sigma_{X}$ & $\sigma_{Y}$ & $\sigma_{Z}$ & $\sigma_{v_{X}}$ & $\sigma_{v_{Y}}$ \\
   		 & kpc & pc & pc & pc & km\,s$^{-1}$ & km\,s$^{-1}$ \\ 
   		\hline
   		CamOB1 & 0.82 & 15 & 9 & 12 & 8 & 8  \\ 
   		CasOB14 & 0.85 & 12 & 9 & 27 & 13 & 5  \\ 
   		CepOB2 & 0.71 & 19 & 8 & 11 & 10 & 6  \\ 
   		CepOB3 & 0.68 & 35 & 14 & 14 & 11 & 6  \\ 
   		CepOB4 & 0.72 & 7 & 16 & 6 & 12 & 7 \\ 
   		Coll140 & 0.35 & 45 & 7 & 28 & 7 & 7  \\ 
   		Coll359 & 0.20 & 22 & 7 & 9 & 8 & 7  \\ 
   		CygOB7 & 0.68 & 14 & 12 & 46 & 11 & 6 \\ 
   		CygOB9 & 0.94 & 15 & 22 & 13 & 10 & 3  \\ 
   		MonOB1 & 0.60 & 3 & 14 & 32 & 7 & 6  \\ 
   		OriOB1 & 0.40 & 14 & 28 & 22 & 4 & 5  \\ 
   		PerOB3 & 0.18 & 23 & 8 & 31 & 12 & 10  \\ 
   		ScoOB2 & 0.15 & 13 & 10 & 12 & 3 & 3  \\ 
   		SctOB3 & 1.25 & 26 & 67 & 45 & 9 & 7  \\ 
   		SgrOB1 & 1.27 & 38 & 87 & 29 & 7 & 4  \\ 
   		Tr27 & 0.90 & 10 & 15 & 18 & 8 & 10  \\ 
   		VelOB2 & 0.38 & 30 & 14 & 42 & 6 & 6  \\ 
   		VulOB4 & 0.80 & 30 & 13 & 16 & 8 & 10  \\ 
   		\hline
   	\end{tabular}
   	\end{center}
   \end{table}
   
  As we are interested in the kinematic properties of the association members, kinematic outliers which are not true association members are potentially a serious problem as they can artificially skew the diagnostics employed in Section 3. For this reason, we impose a velocity cut of $|v_{x}| < 3\sigma_{v_{X}}$  and $|v_y| < 3\sigma_{v_{Y}}$ with $v_x \equiv v_X-\mu_{v_X}$ and  $v_y \equiv v_Y-\mu_{v_Y}$, removing the most extreme velocity sources while maintaining the vast majority of the sample. OB associations were rejected if they did not contain at least 100 stars including at least 10 O- or B-type stars within the 3$\sigma$ limit. This yields a final sample of 18 OB associations. The number of selected members and the number of dispersions in $X$,$Y$ and distance used to select the members of each association are listed in Tables \ref{param_tbl} and \ref{param_tbl_relall}. These kinematic criteria typically filter out less than 1\% of the stars in each association, implying that it has a limited influence on the diagnostics presented in Section 3.

The position-velocity analysis employed in this work depends on a reasonable determination of the centres of the OB associations. The centres of mass of the associations require known masses for all association members, which are not available. Therefore we define two association centres and use both in the subsequent analysis. The first centre is determined to be the mean 4-dimensional ($X$,$Y$,$v_x$,$v_y$) position of the OB stars - this is of course biased towards the most massive and least numerous sources in the association. 
The properties of the OB associations using the mean position and velocity of the OB-type stars are given in Table \ref{param_tbl}.
The second centre is set as the mean 4-dimensional position of all stars in the association, biased towards the more numerous low mass stars. 
Table \ref{param_tbl_relall} shows the properties of the OB associations when using the mean position and velocity of all stars.
Given that one centre is biased towards low mass stars and the other is biased towards high mass stars, it is expected that the centre of mass of each association lies between these two points.
\begin{table*}
	\begin{minipage}{170mm}
		\caption{\label{param_tbl} Table of parameters for all 18 OB associations relative to the centre of OB stars. The first column gives the association names. This is followed by the number of dispersions ($\sigma$) to which association members were selected in 3-dimensional spatial coordinates in order to provide at least 100 stars, at least 10 of which are OB-stars (see Section \ref{TGAS_data}).  The third column contains the number of member stars in each association. Column 4 gives the ratio $N_{v_{r}>0}/N_{v_{r}<0}$ as described in Section 3.1. Columns 5 and 6 give the median radial velocities and radial velocity dispersions. The average velocity uncertainty for each association is given in column 7. The ratio between the radial velocity and the absolute of the tangential velocity (see Section 3.2) and the dispersion in this value are given in the 8th and 9th columns. The final (10th) column displays the radial velocity anisotropy parameter as described in Section 3.3 for each of the associations, where positive (negative) values denote radial (tangential) anisotropy.}
		\begin{center}
			\begin{tabular}{l c c c c c c c c c c}
				\hline
				assoc.  & $N_{\sigma}(XYZ)$ & $N_{\star}$ & $N_{v_{r}>0}/N_{v_{r}<0}$ &  med. $v_{\text{r}}$ (km\,s$^{-1}$) & $\sigma_{v_{r}}$ & $<\Delta v>$  (km\,s$^{-1}$)  & med. $v_{\text{r}} / |v_{\text{t}}|$ & $\sigma_{v_{r} / |v_{\text{t}}|}$ & $\beta$ \\
				\hline
				CamOB1  & 3 & 260 & 1.28 & 1.6 & 10.0 & 4.0& 0.24 & 1.45  & -0.055$\pm$0.006   \\
				CasOB14  & 3 & 134 & 1.39 & 3.1 & 10.7 & 4.2 & 0.50 & 1.47 &   0.043$\pm$0.008    \\
				CepOB2  & 1 & 127 & 1.15 & 1.2 & 8.8 & 4.5 & 0.34 & 2.23 &  0.480$\pm$0.099  \\
				CepOB3   & 1 & 279 & 1.5 & 2.8 & 9.3 & 4.6 & 0.67 &  1.84 &  0.339$\pm$0.054   \\
				CepOB4  & 3 & 388 & 0.90 & -0.4 & 10.0 & 3.6 & -0.07  & 1.19 &  -0.590$\pm$0.070   \\
				Coll140  & 1 & 124 & 1.10 & 0.6 & 9.3 & 2.0 & 0.14 & 1.17 &   -0.868$\pm$0.138 \\
				Coll359  & 3 & 180 & 0.92 & -0.62 & 10.0 & 1.2 & -0.05 & 1.17 &  -0.526$\pm$0.074   \\
				CygOB7  & 1 & 104 & 0.93 & -0.2 & 7.0 & 4.7 & -0.04 & 1.52 &   0.069$\pm$0.017  \\
				CygOB9  & 3 & 314 & 1.15 & 0.5 & 8.0 & 6.7 & 0.09 & 1.34 &   -0.264$\pm$0.036  \\
				MonOB1  & 3 & 126 & 1.10 & 0.5 & 9.8 & 4.8 & 0.07 & 1.56 &   0.219$\pm$0.036 \\
				OriOB1  & 1 & 190 & 0.88 & -0.3 & 3.0 & 1.6 & -0.14 & 1.51  &  0.035$\pm$0.008    \\
				PerOB3  & 3 & 136 & 0.92 & -1.0 & 11.8 & 1.1 & -0.22 & 1.47 &  0.118$\pm$0.024    \\
				ScoOB2  & 3 & 325 & 1.18 & 0.5 & 8.6 & 0.8 & 0.19 & 1.50 &  0.088$\pm$0.013    \\
				SctOB3  & 3 & 100 & 1.13 & 0.4 & 9.9 & 12.7 & 0.04 & 1.44 &  -0.102$\pm$0.021    \\
				SgrOB1  & 3 & 101 & 1.15 & 0.8 & 9.8 & 10.7 & 0.24 & 1.45 &  -0.197$\pm$0.041    \\
				Tr27  & 3 & 117 & 0.92 & -0.9 & 12.3 & 8.2 & -0.10 & 1.80  &  0.166$\pm$0.034    \\
				VelOB2  & 1 & 264 & 0.68 & -1.3 & 6.8  & 2.7 & -0.27 & 1.22  &  -0.769$\pm$0.135    \\
				VulOB4  & 1 & 181 & 0.83 & -1.0 & 10.2 & 4.1  & -0.21  & 1.39 &  -0.168$\pm$0.026    \\
				\hline
			\end{tabular}
		\end{center}
	\end{minipage}
\end{table*}

\begin{table*}
	\begin{minipage}{170mm}
		\caption{\label{param_tbl_relall} Table of parameters for all OB associations relative to the centre of all stars. See also the caption of Table \ref{param_tbl}.}
		\begin{center}
			\begin{tabular}{l c c c c c c c c c}
				\hline
				assoc. &  $N_{\sigma}(XYZ)$ & $N_{\star}$ & $N_{v_{r}>0}/N_{v_{r}<0}$ &  med. $v_{\text{r}}$ (km\,s$^{-1}$) & $\sigma_{v_{r}}$ & $<\Delta v>$  (km\,s$^{-1}$)  & med. $v_{r} / |v_{\text{t}}|$ & $\sigma_{v_{r} / |v_{\text{t}}|}$ & $\beta$ \\
				\hline
				CamOB1		& 3 &	260	&	1.26	&	2.5	&	10.5 & 4.0	&	0.31	&	1.47	&	-0.089$\pm$0.009	\\
				CasOB14		& 3 &	134	&	1.48	&	2.6	&	9.8 & 4.2	&	0.50	&	1.39	&	-0.017$\pm$0.003	\\
				CepOB2		& 3 &	127	&	1.19	&	0.6	&	7.8 & 4.5	&	0.42	&	1.94	&	0.417$\pm$0.081	\\
				CepOB3		& 1 &	279	&	1.47	&	3.1	&	10.3 & 4.6	&	0.71	&	1.93	&	0.369$\pm$0.055	\\
				CepOB4		 & 1 &	388	&	1.10	&	0.7	&	9.5 & 3.6	&	0.08	&	1.16	&	-0.623$\pm$0.076	\\
				Coll140		& 1 &	124	&	1.38	&	1.5	&	8.5 & 2.0	&	0.31	&	1.32	&	-0.796$\pm$0.123	\\
				Coll359		& 3 &	180	&	0.97	&	-0.1	&	9.5 & 1.2	&	-0.01	&	1.26	&	-0.585$\pm$0.082	\\
				CygOB7		& 1 &	104	&	0.79	&	-0.9	&	7.8 & 4.7	&	-0.32	&	1.53	&	0.126$\pm$0.031	\\
				CygOB9		& 3 &	314	&	1.12	&	0.5	&	7.2 & 6.7	&	0.06	&	1.39	&	-0.315$\pm$0.043	\\
				MonOB1		& 3 &	126	&	1.10	&	0.4	&	10.3 & 4.8	&	0.09	&	1.59	&	0.145$\pm$0.021	\\
				OriOB1	 & 1 &	190	&	1.11	&	0.2	&	3.4 & 1.6	&	0.11	&	1.60	&	0.059$\pm$0.014	\\
				PerOB3		& 3 &	136	&	0.86	&	-3.9	&	19.1 & 1.1 &	-0.38	&	1.58	&	0.077$\pm$0.011	\\
				ScoOB2		& 3 &	325	&	1.10	&	0.4	&	9.5 & 0.8	&	0.11	&	1.52	&	0.088$\pm$0.013	\\
				SctOB3		 & 3 &	100	&	1.22	&	0.6	&	11.8 & 12.7	&	0.07	&	1.57	&	-0.014$\pm$0.002	\\
				SgrOB1		 & 3 &	101	&	1.06	&	1.2	&	9.1 & 10.7	&	0.26	&	1.60	&	-0.090$\pm$0.019	\\
				Tr27	 & 3 &	117	&	1.17	&	1.6	&	12.1 & 8.2	&	0.19	&	1.49	&	0.130$\pm$0.025	\\
				VelOB2		& 1 &	264	&	0.80	&	-1.0	&	6.3 & 2.7	&	-0.14	&	1.17	&	-0.739$\pm$0.137	\\
				VulOB4	 & 1 &	181	&	0.97	&	0.0	&	10.0 & 4.1	&	-0.03	&	1.48	&	-0.143$\pm$0.023	\\
				\hline
			\end{tabular}
		\end{center}
	\end{minipage}
\end{table*}

Once the velocities are in the association centre frame, the radial velocities $v_{\text{r}}$ are determined by taking the dot product of the velocity vector with the radial unit vector. The tangential velocity is then obtained by subtraction as $v_{\text{t}}=(v^2-v_{\text{r}}^2)^{1/2}$.
We define a positive radial velocity as moving away from the centre of the OB association and a negative value as moving towards the centre of the association. Similarly, we (arbitrarily) define the tangential velocity as being positive if the motion is anti-clockwise with respect to the previously determined association centre and negative if the motion is clockwise with respect to the association centre.

\subsection{Model distributions}

\begin{figure*}
	\begin{minipage}{170mm}
		\begin{center}
			\includegraphics[width=0.95\linewidth]{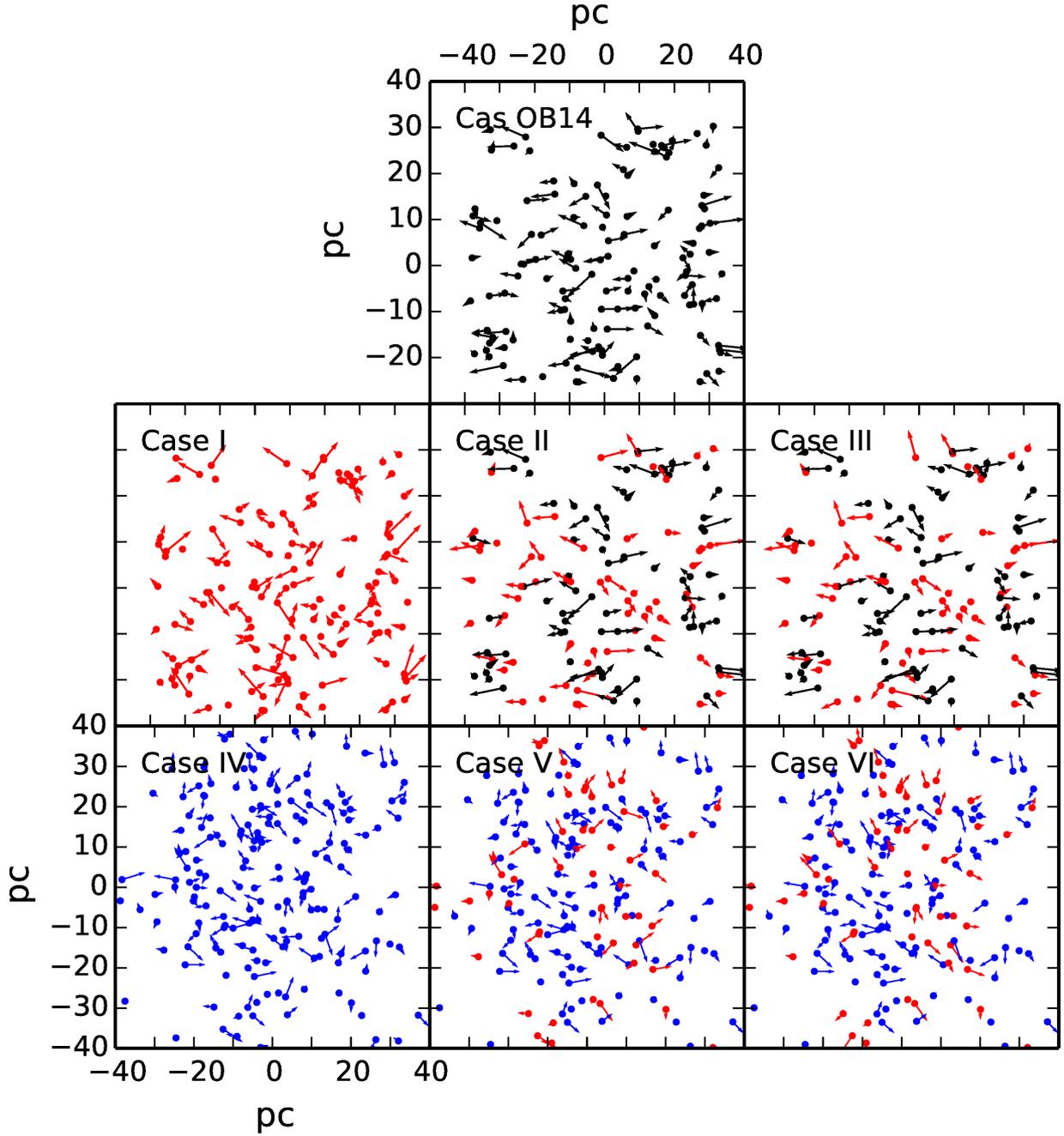}
		\end{center}
		\caption{\label{case_examples} Top: positions of all selected members of Cas OB14 with velocity vectors. Middle row, left to right: Case I model association based on Cas OB14, case II model association based on Cas OB14, and case III model association based on Cas OB14. Bottom row, left to right: Case IV random model association, Case V model association, Case VI model association. Black stars represent original, unaltered position and velocity vectors. Blue stars are those of which the positions and velocities are randomly generated. Red stars are either original or generated stars which have been forced to exhibit randomised directions of motion (case I), or expansion either relative to a local centre (cases II and III) or the association centre (cases V and VI). All positions are in pc. }	
	\end{minipage}
\end{figure*}
In order to assess whether the observed kinematics are consistent with a net radial expansion, we compare our results for the observed OB associations with those of randomly generated velocity distributions.
Six sets of models will be calculated: two distributions representing random velocity fields, two distributions representing localised expansion within associations, and two distributions representing globally expanding velocity fields. These models are therefore representative of no systematic expansion or contraction, expansion from multiple centres within an association (multiply monolithic), and globalised expansion from a single centre (monolithic).
 For reference, we will refer to these six model data sets as cases I, II, III, IV, V, and VI. They are explained below and visualised in Fig. \ref{case_examples}.

The first model set (case I) takes the real positions of the measured OB association data and the magnitude of the motions. The direction of motion is then randomised according to a uniform distribution in azimuthal angle. This process was repeated 20 times to increase the number of renditions yielding a total of 360 associations with randomised directions of motion.

As in the case I models, the case II and III models retain the positions and the velocity magnitudes of the observed OB associations. In case II, the 20 nearest neighbours (in $X$ and $Y$) are selected for each star. We define a local centre as the mean position and velocity of those 20 neighbours, and the radial and tangential components of the stellar velocity are calculated relative to the local centre. One third of the stars within each association are forced to exhibit outwards radial motion with respect to the local centre by taking $v_{\text{r}}$ to be the absolute value of the measured $v_{\text{r}}$. Therefore, this model case represents associations that exhibit localised expansion. In case III we assign the radial velocity to be the absolute of either the radial or tangential velocity components ($v_{\text{r}}=\max{(|v_{\text{r}}|,|v_{\text{t}}|)}$) for the same third of the stars, with the new tangential component being the smaller of the two components. Case III is representative of radially anisotropic localised expansion from multiple centres as is expected from a multiply monolithic formation model.

For the case IV distribution, we use the median and the standard deviations of the position and velocity dispersions of the observed sample of associations to generate a set of random position and velocity dispersions.
These random values were drawn from four Gaussian distributions centred on the median of the measured dispersions in $X$, $Y$, $v_{X}$, and $v_{Y}$ (Table \ref{disp_tbl}) and using the standard deviation of the measured dispersions.
 These were subsequently used to populate 300 model OB associations with random numbers of stars drawn from a uniform distribution occupying the range 100--400.
 A subset of OB stars is defined within each model association by randomly drawing fractions of OB stars from a Gaussian distribution based on the observed median (0.11) and dispersion (0.10) of the OB star fractions in the observed sample.
 Each of the stars is assigned a random position and proper motion, drawn from Gaussian distributions centred on zero using the randomly generated dispersion values.

Cases V and VI are representative of globally expanding configurations of the case IV models.
In case V, one third of the stars within each association are forced to exhibit outwards radial motion with respect to the association centre by taking $v_{\text{r}}$ to be the absolute value of the originally generated $v_{\text{r}}$. On average, this causes $2/3$ of the stars to have positive radial velocities. In case VI, one third of the stars within each association are assigned radial velocities equal to the largest absolute value of the original $v_{\text{r}}$ and $v_{\text{t}}$, i.e.~$v_{\text{r}}=\max{(|v_{\text{r}}|,|v_{\text{t}}|)}$. The tangential velocity is set to the other, smaller component while keeping its original sign. This model ensures that radial motion dominates over tangential motion. This reproduces the strong radial anisotropy ($\beta\sim0.4$) that is expected from a monolithic formation scenario.
\begin{figure*}
	\begin{minipage}{170mm}
		\begin{center}
			\includegraphics[width=0.99\linewidth]{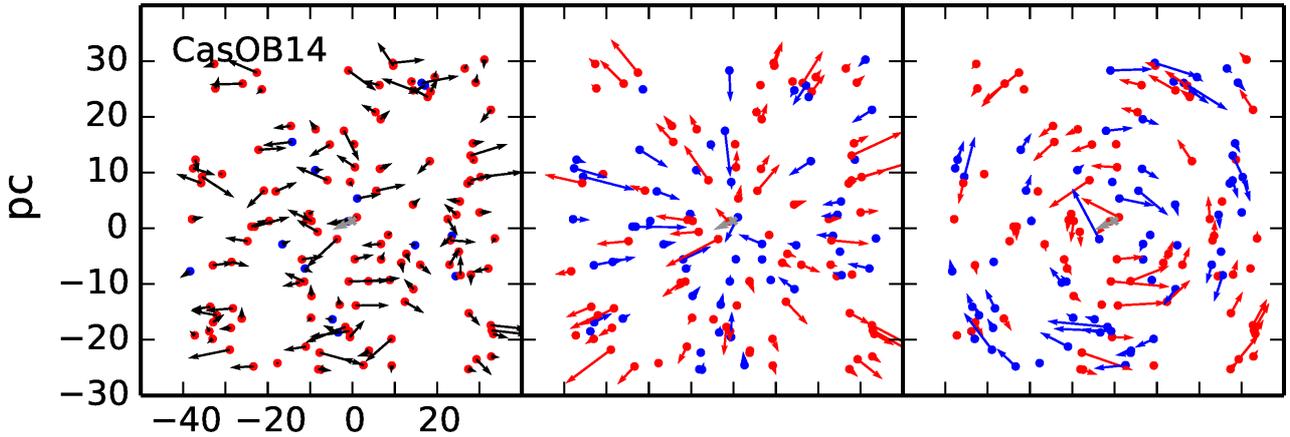}
		\end{center}
		\caption{\label{CasOB_exmpl} X and Y positions (in pc) of all selected members in the OB association Cas OB14 with vectors representing the velocities of all sources relative to the association centre determined using all stars (see text). Left: total velocities. Centre: radial velocity components. Right: tangential velocity components. The blue points in the left panel are OB stars with other stars marked in red. In the centre panel, the points are colour-coded according to $v_{r} < 0$ (blue) and $v_{\text{t}} > 0$ (red). In the right panel, blue points are moving clockwise and red points are moving anti-clockwise with respect to the association centre. In each panel, the relative position and velocity vector of the association based on only OB stars is marked in grey.}
	\end{minipage}
\end{figure*}

The model cases can be summarised as follows:
\begin{enumerate}
	\item Observed stellar positions and absolute velocities with random directions of motion (case I).
	\item Observed stellar positions and absolute velocities where $1/3$ of the stars are guaranteed to have positive radial velocities with respect to the mean position and velocity of the 20 nearest neighbours (case II).
	\item Observed stellar positions and absolute velocities where $1/3$ of the stars are guaranteed to have positive radial velocities and dominant radial motions with respect to the mean position and velocity of the 20 nearest neighbours (case III).
	\item Random stellar positions and velocities drawn from a four-dimensional Gaussian distribution (case IV).
	\item Random stellar positions and velocities drawn from a four-dimensional Gaussian distribution where $1/3$ of the stars are guaranteed to have positive radial velocities (case V).
	\item Random stellar positions and velocities drawn from a four-dimensional Gaussian distribution where $1/3$ of the stars are guaranteed to have positive radial velocities and dominant radial motions (case VI).
\end{enumerate}
Together, these cases span a physically appropriate range of configurations for comparison to the observed kinematic diagnostics discussed below.


\section{Quantifying the kinematics of OB associations}

\label{results_section}

Through analysis of the radial velocities of the member stars of OB associations, it is possible to determine whether the associations are showing signs of strong radial expansion and, by extension, distinguish between the monolithic or hierarchical models of star formation.
The left panel of Figure \ref{CasOB_exmpl} shows the positions and velocities for the OB association Cas OB14. These velocities are then separated into their radial and tangential components in the middle and right panels of Figure \ref{CasOB_exmpl}. These diagrams are shown for all 18 OB associations in Figure \ref{fig:A1}.
In the left panels of Figures \ref{CasOB_exmpl} and \ref{fig:A1}, the OB stars are marked in blue with all other stars marked in red. In the centre panels, the points are colour-coded according to the radial velocity direction: blue for $v_{\text{r}} < 0$ and red for $v_{\text{r}} > 0$. In the right hand panels the points are colour-coded by tangential motion direction: blue for clockwise ($v_{\text{t}} < 0$) and red for anti-clockwise \textquoteleft rotation' ($v_{\text{t}} > 0$).
In all panels, positions and velocities are shown relative to the association centre defined as the mean position and velocity of all stars. The relative position and velocity vector of the centre defined using only the OB stars is marked in grey in each panel.

The general impression of these maps is one of largely stochastic motion, albeit with some substructure in many cases. In the remainder of this section, we employ a number of quantitative tests by considering the cumulative distributions of several key kinematic diagnostics across the sample of OB associations. This allows us to assess the degree to which their kinematics are best described by a net expansion or by random motion.

\subsection{Number ratios}

\begin{figure*}
	\begin{minipage}{165mm}
		\begin{center}
		\includegraphics[width=0.99\linewidth]{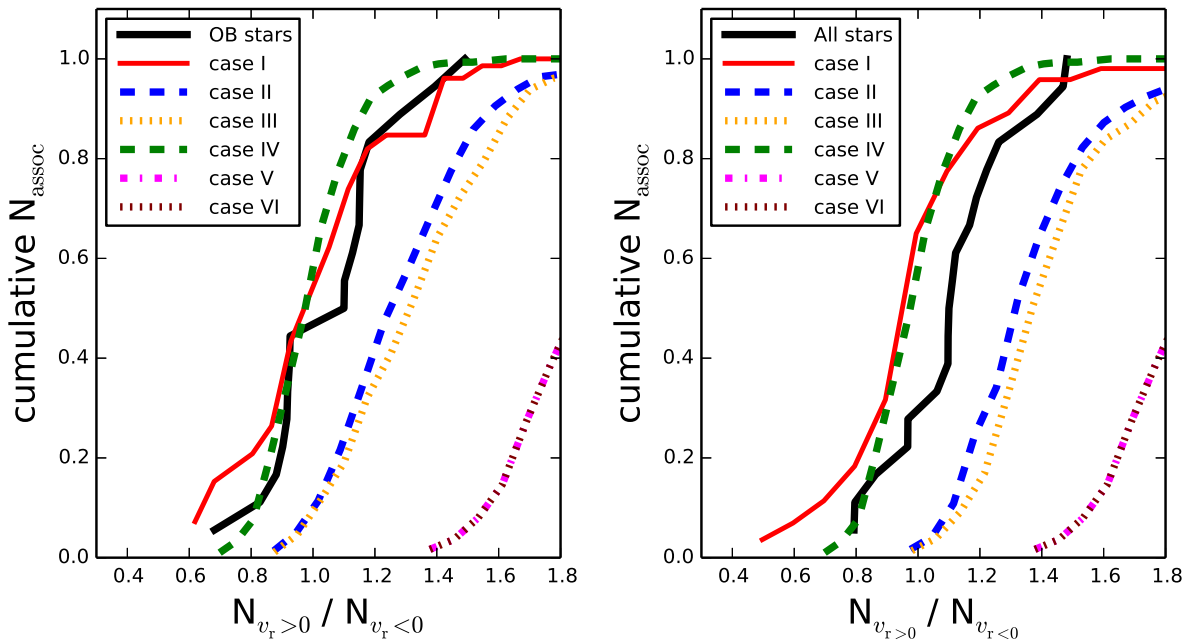}
		\caption{\label{Nrat_fig} Cumulative distribution of the ratio of the number of stars with positive radial velocities over those with negative radial velocities relative to the centre of OB stars (left) and the center of all stars (right). The solid black lines are the measured data for the 18 OB associations analysed in this work. 
			The solid red lines indicate the distributions for the 18 OB associations calculated with real X and Y positions and velocity magnitudes but with randomised directions of motion (case I).
			 Additional lines indicate the case II distribution (blue dashed), case III (orange dotted), case IV (green dashed), case V  (magenta dash-dot) and case VI (maroon dotted) distributions. The case V and case VI distributions are indistinguishable in this figure. 
		The observed OB associations are best reproduced by the random motion models (case I and IV).}
		\end{center}
\end{minipage}
\end{figure*}

If there is a widespread, systematic expansion of present-day OB associations, then it is likely to be evident in the ratio of stars moving away from the association centre to stars moving towards the association centre.
Figure \ref{Nrat_fig} shows the cumulative distribution of the number ratio of  sources with positive (outwards) radial velocities  ($N_{v_{\text{r}}>0}$) to the number of sources with negative (inward) radial velocities ($N_{v_{\text{r}}<0}$) for all OB associations. Also shown in this Figure are the normalised cumulative distributions for each of the six model cases described in the previous section.

It is immediately clear from this figure that, based on number ratios alone, the observed distribution of OB associations are far better represented by purely random distributions of stellar velocities rather than those distributions representing either locally expanding or globally expanding associations.  
There is a departure from the case IV curve tending slightly towards expansion; however, this effect is relatively small and the maximum observed ratio is $\sim$1.5, falling short of what may be expected from a widespread, systematic expansion.
Qualitatively, the case I distributions are the closest approximation to the observed distributions.
When the centres of the OB star populations are used, the case I approach of randomising proper motion directions recreates the departure from the case IV distribution of random positions and velocities. The locally expanding case II and III model distributions fall between the distributions based on random velocity fields and the two globally expanding distributions.
When the centres of the entire stellar populations are used (right panel), the observed systems show a small expansion relative to the random motion models (case I and IV), but this small excess is reproduced by forcing just $\sim$5\% of the stars to exhibit outward radial motion as in cases II, III, V, and VI (rather than the 1/3 used there). This is in strong contrast with the expectation from monolithic models, in which most of the stellar population exhibits expansion \citep{Baumgardt2007}.
Based on these number ratios alone, the observed velocities are certainly consistent with a population of stars with randomly distributed velocities and inconsistent with both global and localised expansion scenarios.

Whilst giving the qualitative impression of a random distribution, the information gained from this simple analysis is limited, providing no indication of the magnitudes of the radial velocities and by itself it is not a definitive measure of the dynamical behaviour of the sample. 
 
\subsection{Median velocity distributions}

To quantify the typical expansion (or contraction) velocities  of the OB associations in our sample, we take the median value of the radial velocities of the stellar populations for each OB association.
The cumulative distributions of the median radial velocities are shown in Figure \ref{PMXp_fig} relative to both the centre of the OB stars and the centre of all stars, along with the six model cases. 
Uncertainties in the observed distributions are calculated by randomly generating 100 cumulative distributions of the median radial velocities by drawing values from a Gaussian distribution for each OB association, using the median radial velocities and the radial velocity dispersions shown in Tables~\ref{param_tbl} and \ref{param_tbl_relall}. These uncertainties are represented by the grey shaded area in Figure \ref{PMXp_fig}.


It is again immediately clear that, although there is a tendency towards positive (outwards) radial velocities, the observed distribution falls closer to the case I (randomised velocity directions) and case IV (randomised positions and velocities) distributions than to any of the expanding models, regardless of which association centre is used. 
Within the uncertainties of the observed distributions (represented by the grey shaded area in Figure \ref{PMXp_fig}), the lower end of the observed distributions are consistent with the case IV random distribution. However, the case I distribution provides a closer match than the case IV distribution over the positive radial velocity region of  parameter space.
The extreme positive ends of the distributions partially overlap with the case II, III, and V model distributions; however, the positive parts of the distributions are far more consistent with the case I distribution overall.
Therefore, the tendency towards positive values could be a result of positional substructure in the apparently high $v_{\text{r}}$ associations rather than a genuine radial velocity effect, as the case I distribution retains the original geometry of the observed associations, and therefore any existing positional substructure. 

\begin{figure*}
	\begin{minipage}{165mm}
		\begin{center}
			\includegraphics[width=0.99\linewidth]{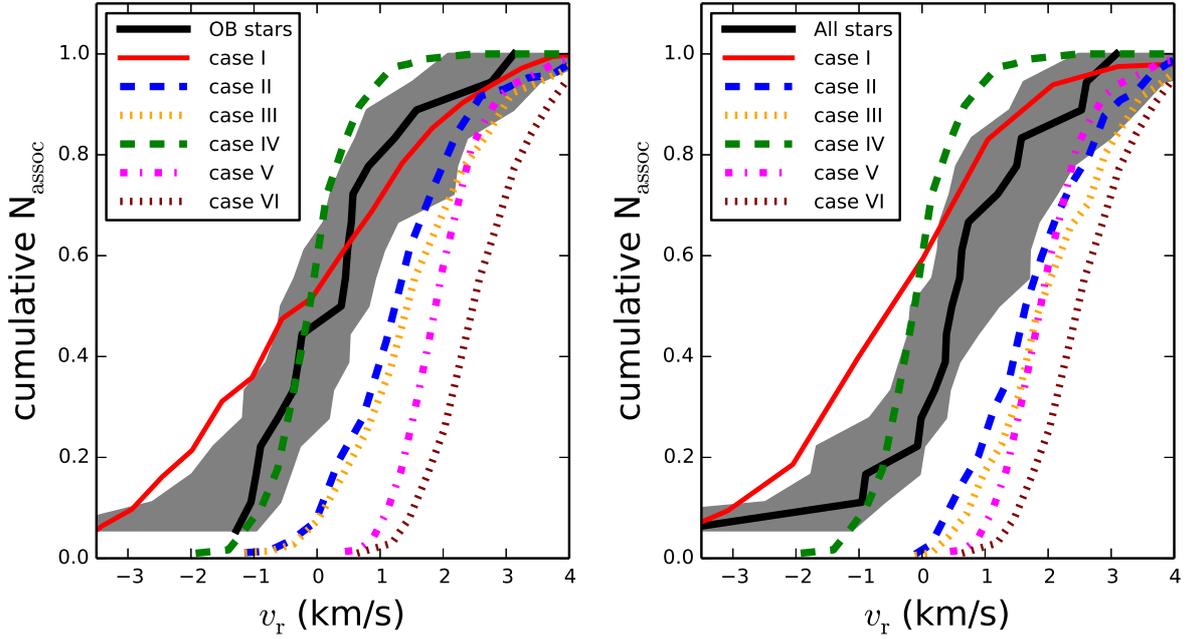}
			\caption{\label{PMXp_fig} Cumulative distributions showing the median radial velocities for all associations relative to the centre of OB stars (left) and the centre of all stars (right). 
				The solid black lines are the measured data for the 18 OB associations analysed in this work. Uncertainties in the observed distributions are represented by the grey shaded area.
				The solid red lines indicate the distributions for the 18 OB associations calculated with real X and Y positions and velocity magnitudes but with randomised directions of motion (case I).
			Additional lines indicate the case II distribution (blue dashed), case III (orange dotted), case IV (green dashed), case V  (magenta dash-dot) and case VI (maroon dotted) distributions. The observed OB associations are best reproduced by a combination of the random motion models (cases I and IV).}
		\end{center}
	\end{minipage}
\end{figure*}

Overall, the median radial velocities in Figure \ref{PMXp_fig} are much smaller than the radial velocity dispersions (see Tables \ref{param_tbl} and \ref{param_tbl_relall}). This indicates that any expansion of the associations 
 is unlikely to have resulted from an impulsive event (such as gas expulsion) that caused a possible progenitor cluster to become unbound. Instead, these radial velocities are more consistent with the gradual, secular evolution of the nascent velocity field, as inherited from the turbulent ISM.
 
 In Tables \ref{param_tbl} and \ref{param_tbl_relall}, the seventh column shows the mean velocity uncertainty of the members of each association. In two of the 18 OB associations (Sct OB3 and Sgr OB1), the radial velocity dispersions are lower than the average uncertainties. These two associations are at relatively high distances where the velocity uncertainties are dominated by the uncertainties in parallax in the TGAS catalogue.The TGAS parallax uncertainties are likely to be overestimated (see the discussion in Section 4.2) which can account for the measured velocity dispersions being smaller than the average uncertainties.
 
While the median velocity distributions convincingly rule out systematic rapid ($>3$\,km\,s$^{-1}$) expansion, they fail to take into account any information regarding the tangential component of the stellar velocities.
In Figure \ref{PMXp_PMYp_fig}, the median is taken of the ratio of the radial velocity to the absolute value of the tangential velocity ($v_{\text{r}} / |v_{\text{t}}|$) across all stars in the OB associations. This probes the expansion rates in units of the typical velocity in the tangential direction.
When the tangential velocities are taken into account, the case VI distribution is clearly distinguished from the other expanding distributions (cases II, III, and V), indicating that this is a powerful diagnostic of global, strongly anisotropic expansion. 

The overall trends of Figures \ref{PMXp_fig} and \ref{PMXp_PMYp_fig} are very similar. However, the normalised expansion velocities of Figure \ref{PMXp_PMYp_fig} provide a better match to cases I and IV than the absolute expansion velocities in Figure \ref{PMXp_fig}. The upper end of the observed distribution relative to the centre of all stars falls between the randomly distributed case IV and the locally expanding case II distributions but is well characterised by the case I distribution. At its most extreme, the observed distribution is consistent with both locally expanding model distributions. However, this only concerns up to three OB associations out of the 18 analysed in this work.
The radially anisotropic, globally expanding case VI distribution remains entirely inconsistent with observations.

\begin{figure*}
	\begin{minipage}{165mm}
		\begin{center}
			\includegraphics[width=0.99\linewidth]{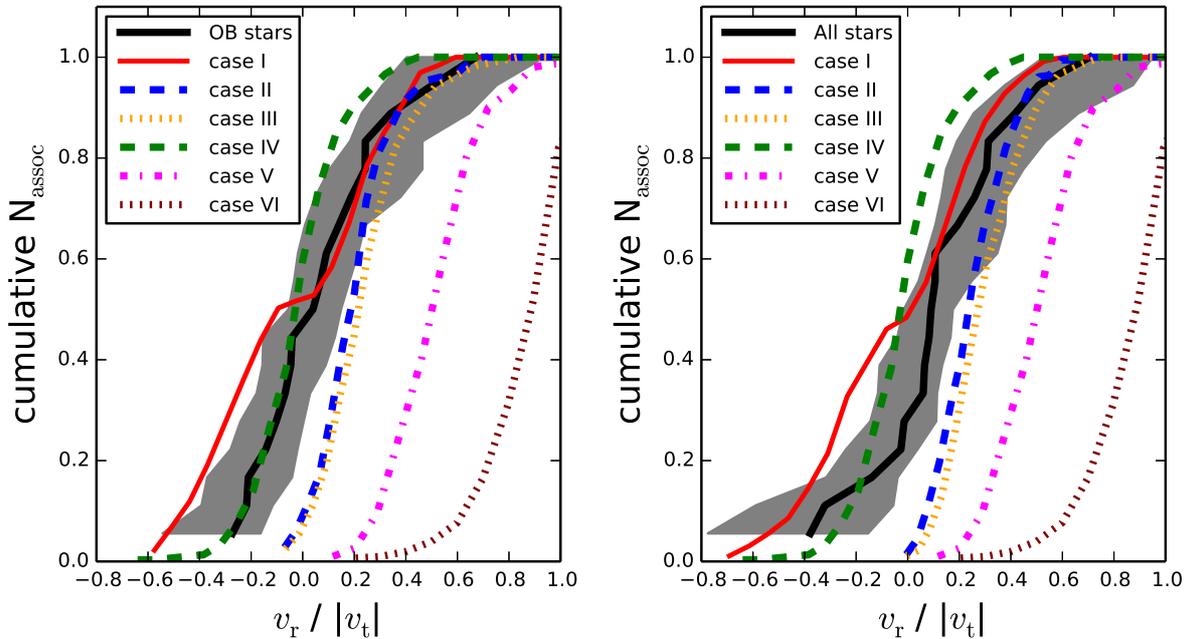}
			\caption{\label{PMXp_PMYp_fig} Cumulative distributions showing the median $v_{\text{r}}$ / $| v_{\text{t}} |$ values for all associations relative to the centre of OB stars (left) and the center of all stars (right). Lines and symbols have the same meaning as for Figure \ref{PMXp_fig}. Again, the observed OB associations are best reproduced by a combination of the random motion models (cases I and IV).}
		\end{center}
	\end{minipage}
\end{figure*}

\subsection{Anisotropy}

The radial anisotropy of a system represents the degree to which the radial velocity component dominates over the tangential velocity component. A radially anisotropic velocity field, retained for many initial crossing times (or tens of Myr), is a firm prediction of the monolithic star formation model.
The 2-dimensional anisotropy parameter ($\beta$) determined for the OB associations in this work is directly analogous to the 3-dimensional parameter and takes the form:
\begin{equation}
\label{anieqn}
\beta = 1-\frac{<v_{\text{t}}^{2}>}{<v_{\text{r}}^{2}>} \text{.}
\end{equation}
While the three-dimensional form carries an additional factor of 2 in the denominator due to the double dimensionality of the tangential component, this is omitted when using a 2-dimensional space.  A value of zero indicates that the radial and tangential velocity components are equal, while negative values and positive values indicate dominant tangential and radial components, respectively. In a system experiencing a period of systematic expansion (or contraction), $\beta$ is expected to have a positive value. This is a critical quantity for answering whether OB associations are expanded clusters, because $N$-body simulations show that the anisotropy remains systematically high ($\beta > 0.2$) for long periods (tens to hundreds of initial crossing times, or tens of Myr) after gas expulsion \citep{Baumgardt2007}.

The cumulative distributions for the anisotropy parameter are shown in Figure \ref{beta_fig}, using both the centre of OB stars and the centre of all stars.
The model case where we take the absolute $v_{\text{r}}$ value for 1/3 of a set of randomly generated sources (case V) does not affect the anisotropy parameter, because it enters squared in equation (\ref{anieqn}). Only the case III and VI distributions represent strongly radially anisotropic expansion. 

 As for the previous quantities, the case I distribution fits well to observed data in the positive region of parameter space, whereas the case II, III, IV and V distributions provide a far better fit to the data at negative anisotropies. 
Qualitatively, the observed cumulative distributions of $\beta$ are well matched by the case I, II, III, IV, and V distributions, while they are entirely inconsistent with case VI (where the radial velocity component is dominant). 
The case III anisotropic localised expansion models  fall very close to the case II non-anisotropic locally expanding models; however, there is always a slight anisotropy excess in the case III models with respect to the case II models.
The median of the anisotropy parameters of the OB associations in this sample is approximately zero. The radial component is therefore certainly not dominant, indicating no systematic tendency towards expansion or contraction. Likewise, there is no observed bias towards tangential motion, which implies that the observed stars are not moving in well ordered orbits around a common centre of mass. Therefore, it is clear that, even if none of the model cases presented here fit the observed data well, the observed population of OB associations cannot be produced by singularly monolithic expansion. However, expansion from multiple centres cannot be ruled out by the anisotropy parameter alone. Multiply monolithic cluster formation is ruled out by the combination of quantities discussed previously.

\begin{figure*}
	\begin{minipage}{165mm}
		\begin{center}
			\includegraphics[width=0.99\linewidth]{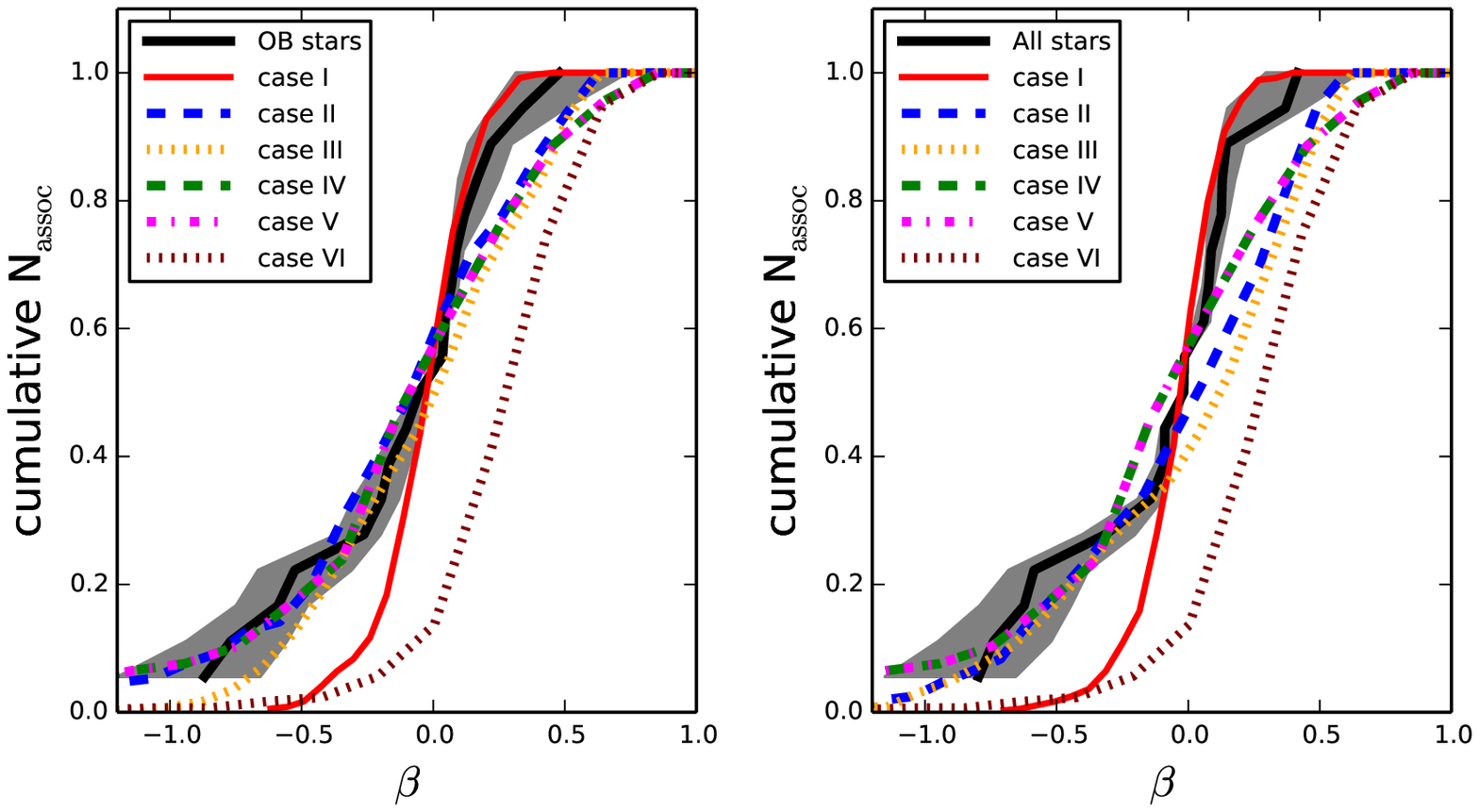}
			\caption{\label{beta_fig} Cumulative distributions of the anisotropy parameter $\beta$, for all associations relative to the centre of OB stars (left) and the center of all stars (right). Lines and symbols have the same meaning as for Figure \ref{PMXp_fig}. Again, the observed OB associations are best reproduced by a combination of the random motion models (cases I and IV). The anisotropy does not allow a distinction between case IV and case V, but case VI expansion is firmly ruled out.}
		\end{center}
	\end{minipage}
\end{figure*}

\subsection{Kolmogorov-Smirnov tests}
\begin{table*}
	\begin{minipage}{170mm}
		\caption{\label{KS_table} Kolmogorov-Smirnov statistics and associated {\it p}-values calculated using the centre of the OB star members for each association. The quantity for which the test is carried out is given in the first column. The following columns show the KS statistic and the corresponding {\it p}-values for each of the model cases. 
			Case I uses the measured positions and absolute velocities of the stars in the 18 measured OB associations, but with randomised proper motion directions.
			Case IV  uses randomly generated model associations, with each following a Gaussian distribution in X,Y,$v_x$ and $v_y$. In case V one third of the stars in the case IV associations are forced to be expanding by using the absolute value of $v_{\text{r}}$. In case VI, the velocity vectors of one third of the stars were changed such that the radial velocity is set by the largest absolute value of $v_{\text{r}}$ and $v_{\text{t}}$, ensuring that radial motion is dominant. The final column shows the KS test results for when the case I distribution is used above the threshold for expansion (1 for $N_{v_{\text{r}}>0}/N_{v_{\text{r}}<0}$ and 0 for all other parameters) and the case IV distribution is used at values lower than the threshold.}
		\begin{center}
			\begin{tabular}{l c c c c c c c c c c c c c c}
				\hline
				&		\multicolumn{2}{c}{case I} & \multicolumn{2}{c}{case II} & \multicolumn{2}{c}{case III} & \multicolumn{2}{c}{case IV} & \multicolumn{2}{c}{case V} & \multicolumn{2}{c}{case VI} & \multicolumn{2}{c}{case I $+$ case IV} \\
				Quantity	&	KS	& $p$ & KS & $p$ & KS	& $p$ & KS & $p$ & KS & $p$ & KS	& $p$ & KS	& $p$\\
				\hline
				$N_{v_{\text{r}}>0}/N_{v_{\text{r}}<0}$ & 0.239 & 0.245 & 0.519  & 0.000  & 0.583 & 0.000 & 0.316 & 0.053 & 0.963 & 0.000 & 0.963 & 0.000 & 0.244 & 0.227 \\
				$v_{\text{r}}$ & 0.306 & 0.065 & 0.525 & 0.000 & 0.572 & 0.000 & 0.312 & 0.057 & 0.758 & 0.000 & 0.816 & 0.000 & 0.146 & 0.831  \\
				$v_{\text{r}}/ |v_{\text{t}}|$ & 0.272 & 0.131 & 0.439 & 0.002 & 0.475 & 0.001 & 0.212 & 0.385 & 0.807 & 0.000 & 0.921 & 0.000 & 0.112 & 0.977 \\
				$\beta$ & 0.244 & 0.222  & 0.156 & 0.765 & 0.231 & 0.283 & 0.231 & 0.303 & 0.231 & 0.303 & 0.753 & 0.000 & 0.169 & 0.681  \\
				mean & 0.265 & 0.166 & 0.410 & 0.192 & 0.465 & 0.071 & 0.268 & 0.200 & 0.690 & 0.076 & 0.863 & 0.000 & 0.168 & 0.679 \\
				\hline
			\end{tabular}
		\end{center}
	\end{minipage}
\end{table*}
\begin{table*}
	\begin{minipage}{170mm}
		\caption{\label{KS_table_relall} KS statistics and associated {\it p}-values calculated using the centre of all stars for each association. See also the caption of Table \ref{KS_table}.}
		\begin{center}
			\begin{tabular}{l c c c c c c c c c c c c c c}
				\hline
				&		\multicolumn{2}{c}{case I} & \multicolumn{2}{c}{case II} & \multicolumn{2}{c}{case III} & \multicolumn{2}{c}{case IV} & \multicolumn{2}{c}{case V} & \multicolumn{2}{c}{case VI} & \multicolumn{2}{c}{case I $+$ case IV} \\
				Quantity	&	KS	& $p$ & KS & $p$ & KS	& $p$ & KS & $p$ & KS & $p$ & KS & $p$ & KS	& $p$ \\
				\hline
				$N_{v_{\text{r}}>0}/N_{v_{\text{r}}<0}$ & 0.414 & 0.004 & 0.667 & 0.000 & 0.739 & 0.000 & 0.420 & 0.003 & 0.970 & 0.000 & 0.970 & 0.000 & 0.242 & 0.235  \\
				$v_{\text{r}}$ & 0.303 & 0.069  & 0.625 & 0.000 & 0.689 & 0.000 & 0.423 & 0.003  & 0.650 & 0.000 & 0.760 & 0.000 & 0.308 & 0.063  \\
				$v_{\text{r}}/ |v_{\text{t}}|$ & 0.297 & 0.078 & 0.517 & 0.000 & 0.578 & 0.000 & 0.389 & 0.008 & 0.770 & 0.000& 0.921 & 0.000 & 0.295 & 0.084 \\
				$\beta$ & 0.231 & 0.283 & 0.281 & 0.111 & 0.353  & 0.021  & 0.280 & 0.126 & 0.280 & 0.126 & 0.813 & 0.000 & 0.197 & 0.487  \\
				mean & 0.311 & 0.109 & 0.523 & 0.028 & 0.590 & 0.005 & 0.378 & 0.035 & 0.668 & 0.032 & 0.866 & 0.000 & 0.261 & 0.217 \\
				\hline
			\end{tabular}
		\end{center}
	\end{minipage}
\end{table*}

After qualitatively comparing the kinematic properties of the observed OB associations to our model distributions, we now quantify this comparison further.
In order to assess the likelihood that model and observations are drawn from the same parent samples, we carry out two-sample Kolmogorov--Smirnov (KS) tests between the observed OB association data and each of the model case distributions.
The KS tests return a KS statistic which is a normalised maximum distance between the two cumulative distributions and a {\it p}-value which represents the likelihood that the two distributions were drawn from the same parent sample.
 The resulting KS statistics and corresponding {\it p}-values are given in Tables \ref{KS_table} and \ref{KS_table_relall} using the centres of OB stars and the centres of all stars, respectively.

For the number ratio of sources with positive versus negative values of the radial velocity ($N_{v_{\text{r}}>0} / N_{v_{\text{r}}<0}$) relative to the OB star centre, the case I distribution has the lowest KS statistic (0.239) alongside the highest {\it p}-value (0.245). While neither the KS statistic, nor the {\it p}-value indicate a particularly good fit, this indicates that it is the most likely model from which the observed data were drawn out of those tested. The case I distribution and the observed distribution relative to the centre of all stars do not agree as well, with a KS statistic and {\it p}-value of 0.414 and 0.004, respectively. Regardless of which centre is used, the case I distribution is the closest match to the observations and both of the globally expanding configurations (cases V and VI) are highly inconsistent with the observed distribution, exhibiting KS statistics of 0.96 and 0.97 as well as extremely low {\it p}-values. The locally expanding model distributions (cases II and III) are also inconsistent with observations, with KS statistics in the range $0.5-0.8$ and extremely low {\it p}-values. The case IV distribution lies between the case I and the expanding distributions in terms of both KS statistics and {\it p}-values.

For the median radial velocity distributions, the case I distribution is again the best fit to the observed data, with relatively low KS values of $\sim$0.3 and {\it p}-values of $\sim$0.07 regardless of which centre is used. While case IV does not provide as good a fit as case I, it represents a significant improvement on all of the expanding model cases in both KS statistic and {\it p}-value. 
While none of the model distributions provide particularly good fits to the observed data, with {\it p}-values of $\sim 0$, all of the expanding model cases are ruled out based on the observed median velocity distributions, regardless of whether the expansion is global or localised.
When the median radial velocities are normalised by the tangential velocities, the KS statistics again suggest that case I and IV are the best fits to the data, with cases V and VI being strongly inconsistent with the observations. 

For the anisotropy parameter, the case I, II, IV and V models provide similar fits to the observed distribution, consistently providing KS statistics below 0.3 and {\it p}-values greater than 0.1. Case III exhibits similar KS statistics and {\it p}-values relative to the centre of OB stars but the {\it p}-value is significantly lower (0.021) when the centre is defined as the mean position and velocity of all stars. 
Meanwhile, the strongly anisotropic globally expanding case VI results in high KS statistics (0.753 and 0.813) and extremely low {\it p}-values, indicating that case VI is inconsistent with the observed distributions. While none of the model distributions fit the observed data well, only the globally expanding and anisotropic case VI model distribution is conclusively ruled out by the anisotropy parameter.

The mean {\it p}-values of the six model cases in Tables~\ref{KS_table} and~\ref{KS_table_relall} summarise the above discussion, in that case I and IV best reproduce the observed kinematics.\footnote{The mean value of case II, when using the centre of OB stars, is also quite high but this is owing only to the anisotropy parameter.} The globally expanding models of case V and VI are firmly ruled out by the observations. While the locally expanding models are not ruled out by anisotropy, they are firmly ruled out by the remaining three parameters. 
Regardless of the parameter being tested, or whether the centre of the OB stars or the centre of all stars is used as the reference frame, the cases that represent simple random distributions (case I and case IV) yield lower KS statistics than the globally expanding scenarios (case V and case VI). This also holds in the case of localised expansion from multiple points (case II and III), except for the anisotropy parameter. This is in good agreement with the qualitative interpretations of Figures \ref{Nrat_fig}--\ref{beta_fig}. 
In general, the sample of OB associations presented in this work are therefore best described as close to random velocity fields with no evidence of systematic expansion or contraction.

It is clear from Figures \ref{Nrat_fig}--\ref{beta_fig} that a combination of cases I (at high, expansion-like values) and IV (at low, expansion-averse values) will qualitatively provide the best fit to the data in every case.
 In the final columns of Tables \ref{KS_table} and \ref{KS_table_relall}, we present KS test results for a combination of the case I and case IV models where the values from case I are used above the threshold value between expansion and contraction (1 for the $N_{v_{\text{r}}>0}/N_{v_{\text{r}}<0}$ ratio and 0 for all other parameters), and case IV below that threshold.
 When these two distributions are combined in such a manner, it provides the lowest KS statistics and highest {\it p}-values for all parameters except for $N_{v_{\text{r}}>0}/N_{v_{\text{r}}<0}$  when the centre is defined as the mean position and velocity of OB-type stars. The case I models retain the original geometries of the observed associations. It is therefore plausible that the departure of the observed distribution from the purely random case IV distribution towards expanding velocity fields is due to a geometric effect such as positional substructure.

\section{Discussion}

In this paper, we have quantified the kinematics of the members of 18 OB associations with high-quality astrometry from the {\it Gaia}-TGAS catalogue of stellar parallaxes and proper motions. The position--position figures with velocity vectors shown in Figure \ref{fig:A1} show that there are no immediately obvious trends in velocity fields or in the radial or tangential components of these velocities. The degree to which these associations can be described as being expanding has been quantified using four key kinematic diagnostics, finding no evidence of systematic rapid expansion of nearby OB associations.
In conclusion, these results show that the observed kinematics of nearby OB associations are inconsistent with the gas-expulsion driven expansion of the singularly and multiply monolithic cluster formation scenarios. While hierarchical star formation is neither ruled out nor confirmed by this work, the elimination of monolithic cluster formation scenarios must favour hierarchical star formation.

\subsection{Implications of this work}

We have used the radial  and tangential velocity components in the plane of the sky ($v_{\text{r}}$ and $v_{\text{t}}$, respectively) of the association members to quantify four kinematic properties for each of the associations: the stellar number ratio $N_{v_{\text{r}}>0}/N_{v_{\text{r}}<0}$, the median radial velocity ($v_{\text{r}}$),  the median radial velocity normalised by the tangential velocity ($v_{\text{r}}/|v_{\text{t}}|$), and the radial anisotropy parameter ($\beta$). 
The cumulative distribution of each of these values over all OB associations is tested against model distributions generated using randomised velocity distributions and expanding distributions (both from a single centre and from multiple centres).
 In each of these tests, the measured sample lies much closer to the random velocity field distributions than the distributions resulting from the globally expanding models. The localised expansion models are ruled out in all parameters except for the anisotropy parameter. This is quantified through the use of KS tests performed between each model distribution and its observed counterpart, shown in Tables \ref{KS_table} and \ref{KS_table_relall}. While none of the models provide a particularly good fit to the observed data, it is clear from these tests that the observed distributions in radial velocity and anisotropy are inconsistent with a  scenario in which OB associations exhibit any significant net expansion due to having been much more compact in the past. Even though some OB associations do show signs of expansion, these can easily be reconciled with a random velocity distribution when the original association geometries are retained. This is most likely the result of substructure in the $XY$ plane, as this is  the only difference between the case I and case IV distributions.  There is no widespread, systematic expansion of present-day OB associations.

\citet{Baumgardt2007} find that in the event of early cluster disruption resulting from rapid gas expulsion, stars acquire strongly radially anisotropic velocity fields. This is not seen in the sample of OB associations presented here.
Indeed, no evidence has been found to support the view that OB associations are undergoing dispersal by radial expansion following gas expulsion. 
The consistency of the observed velocity fields with random velocity distributions (within the uncertainties) indicates that there is no systematic expansion or contraction of OB associations. 
This is inconsistent with a picture of star formation in which high density, gravitationally bound clusters are a fundamental unit of star formation and subsequently expand in response to gas expulsion. 
This sample of associations is therefore far more consistent with a hierarchical model of star formation than singularly or multiply monolithic star formation models.
 From this, and in light of previous studies of OB associations (e.g. \citealt{Wright2014,Wright2016}), we draw the conclusion that it is unlikely that OB associations are the expanded remnants of dense clusters and that therefore not all stars form in high-density clustered environments. Of the two widely adopted models of association formation (monolithic and hierarchical), our findings favour a hierarchical star formation scenario, in which stars are formed over a scale-free, hierarchically structured continuum of environments. This continuum allows for the in-situ formation of the full range of young stellar populations, from dense clusters at the high-density end, through OB associations at intermediate densities, to relatively isolated star formation at low densities.

\subsection{Caveats and other considerations}

 Following the release of the TGAS catalogue, the appropriateness of the large parallax uncertainties presented in the catalogue has been called into question.
 The parallax uncertainties in the TGAS catalogue have been inflated with respect to the internally derived formal uncertainties \citep{Lindegren2016} by:
\begin{equation}
  \begin{aligned}
\sigma_{\text{TGAS}}(\pi) = \sqrt{[A\sigma_{int}(\pi)]^{2}+\sigma_{0}^{2}} \text{\,\,} \text{\, , \,} (A,\sigma_{0}) = (1.4,0.20 \text{\,mas}) \text{.}
  \end{aligned}
\end{equation}
Since the release of the catalogue, a number of studies have shown that these uncertainties (which are conservative by design) are significantly overestimated. By comparing the TGAS parallax measurements with those determined from the RR Lyrae period-luminosity relation, \citet{Gould2016} found that the uncertainties in TGAS are overestimated and suggest that more appropriate values would be $(A,\sigma_{0}) = (1.1,0.10 \text{\,mas})$. By contrast, when comparing the TGAS parallaxes to those derived from the RAVE, \citet{McMillan2017} find that the random uncertainties in the TGAS catalogue are overestimated by $\sim$0.2~mas. 
We choose to use the original, conservative uncertainties, because the disagreement between the Gould and McMillan papers shows that the extent to which the uncertainties are overestimated is not uniquely characterised. The overestimation of the parallax uncertainties can account for the small radial velocity dispersions compared to the average velocity uncertainties of the two most distant OB associations in this study  (see Section 3.2).

 In addition, \citet{McMillan2017} show that for a small region of the sky, the parallaxes in the TGAS catalogue are systematically higher than those determined by the Radial Velocity Experiment (RAVE, \citealt{Steinmetz2006}) by up to 0.4\,mas. At the distance of 1\,kpc this would mean that the distances in the TGAS catalogue could be underestimated by up to $\sim$300\,pc. The affected region lies at approximately $\lambda\sim0$\degree\,and $\beta\sim-60$\degree\,in ecliptic coordinates. We have verified that none of the OB associations in our sample lie in a conservatively-chosen region between $-30$\degree and $30$\degree\,in ecliptic longitude and lower than $-20$\,\degree\,in ecliptic latitude. Therefore our results are unaffected by this discrepancy in parallax.

Substructure and mass segregation are powerful diagnostics of the formation and past evolution of associations, which in turn can be key in distinguishing between different models of star formation \citep{Parker2014,Wright2014}. While we speculate that positional substructure retained in the case I models may explain the better fit to observational data than the purely random case IV models, kinematic and positional substructure has not been directly addressed in this study. This is because the limited number of OB association members that have been selected makes any detailed analysis of such small structures highly uncertain. It is anticipated that future {\it Gaia} data releases will allow high precision, kinematic studies of individual substructures and clusters within OB associations.

\citet{Melnik2017} have also made use of the TGAS catalogue to study the kinematics of OB associations, finding that their sample of observed OB associations are globally unbound. Their result does not fundamentally conflict with the results of this paper. An obvious potential point of contention is the high expansion velocities found for Per OB1 and Car OB1 of $6.3~{\rm km}~{\rm s}^{-1}$. These associations were not deemed to contain a sufficient number of stars for our study after the membership selection criteria imposed in Section 2, so we cannot quantitatively comment on these velocities directly. However, our findings do suggest that random velocity fields can easily reproduce apparent median expansion velocities of at least 3\,km\,s$^{-1}$ when original association geometries are retained, including any possible substructure. 
It should also be noted that \citet{Melnik2017} find velocity dispersions of the order of 4\,km\,s$^{-1}$, around half the average radial velocity dispersion of our larger sample of association members (9\,km\,s$^{-1}$). This could be related to the smaller samples of association members used in \citet{Melnik2017}, making all averages and dispersion measurements more susceptible to outliers.
In view of the above considerations the finding of expanding OB associations from \citet{Melnik2017} should be treated with some caution, because the physical interpretation of the adopted metrics is not unambiguous.

The sample of associations analysed in this study is not complete, but it does represent the subset that can be studied up to the maximum considered distance of 1.6\,kpc at the desirable precision using the available data from DR1. Moreover, due to the relatively high uncertainties in distances derived from TGAS parallaxes (compared to the upcoming {\it Gaia} DR2), it is highly likely that the membership selection is both incomplete and has a number of contaminants. 
The upcoming {\it Gaia} data releases will allow a study of parallaxes and proper motions for OB association members derived entirely from {\it Gaia} data, allowing for a significant improvement in precision. This will allow for an unbiased approach to selecting OB associations without the prior selection criterion of an OB association appearing in the literature. Moreover, the improved precision of distances derived from parallaxes in future {\it Gaia} data releases will increase the maximum distance to which the techniques employed in this work can be applied, and increase the number of successfully selected association members.
With the combination of {\it Gaia} DR2 and line-of-sight velocities from large scale radial velocity surveys, we anticipate a future  high precision six-dimensional study of a significantly larger sample than of the OB associations presented here.

\section{Conclusions}

In this paper, we quantify the kinematics of 18 nearby OB associations using the {\it Gaia} TGAS catalogue for which we could identify at least 100 stellar members. The goal of this analysis is to test whether these associations are the relics of single or multiple expanded compact clusters. We measure four key parameters:  the stellar number ratio $N_{v_{\text{r}}>0}/N_{v_{\text{r}}<0}$, the median radial velocity ($v_{\text{r}}$),  the median radial velocity normalised by the tangential velocity ($v_{\text{r}}/|v_{\text{t}}|$), and the radial anisotropy parameter ($\beta$). We compare the observed cumulative distributions of these parameters with those of model OB associations that are representative of random velocity fields, expansion from a single point, and expansion from multiple points. A systematic global expansion of OB associations from a single point, as would be expected from a singularly monolithic cluster formation scenario, is firmly ruled out by each of the four parameters measured in this study. This is the case regardless of whether the expansion is anisotropic or non-anisotropic. Likewise, a multiply monolithic cluster formation model, in which expansion occurs from many points, is ruled out by the number ratio, median radial velocity, and normalised median radial velocity parameters. Within the uncertainties, the observed associations are largely consistent with the model distributions that represent randomised motions of stars with no systematic expansion or contraction. Where small net expansions ($<3$\,km\,s$^{-1}$) are present, these can easily be reconciled with random velocity fields when retaining the geometric configurations of the observed associations, which preserve any positional substructure. 
	
	Based on the above results, we infer that OB associations most likely formed as large-scale systems, with hierarchical structures similar to those of molecular clouds. Therefore, we must conclude that only a subset of stars form in gravitationally-bound clusters.
	Given that, in the solar neighbourhood, most stars older than a few Myr reside in associations \citep{Blaauw1964,LadaLada2003}, this implies that a minority of all stars form in bound clusters under solar neighbourhood conditions.


The perspectives for the continuation of our analysis using the diagnostic power of future {\it Gaia} releases are highly encouraging. By using only 18 OB associations from the first Gaia data release, the present study already conclusively rules out both the singularly and multiply monolithic cluster formation scenarios where all OB associations are the expanded relics of either single or multiple dense clusters following their expansion due to gas expulsion. Instead, it favours a model in which OB associations originate in-situ and inherit the structure and dynamics of the hierarchical ISM. Given that many more stars are found in OB associations than in dense clusters (e.g. \citealt{LadaLada2003}), this immediately implies that most stars do not form in clusters in the solar neighbourhood. While this work has expanded the handful of OB associations studied this way in the literature to about 20, future Gaia releases will enable systematic surveys of hundreds of OB associations. The next couple of years  promise to mark great progress in our understanding of spatially clustered star formation.

\section*{Acknowledgements}

JLW and JMDK acknowledge support from Sonderforschungsbereich SFB 881 \textquotedblleft The Milky Way System\textquotedblright\, (subproject P1) of the German Research Foundation (DFG).
JMDK gratefully acknowledges funding from the DFG in the form of an Emmy Noether Research Group (grant number KR4801/1-1, PI Kruijssen), from the European Research Council (ERC) under the European Union's Horizon 2020 research and innovation programme via the ERC Starting Grant MUSTANG (grant agreement number 714907, PI Kruijssen). JLW and JMDK are grateful to the participants of the SFB881 Workshop in Kloster Sch\"{o}ntal in April 2017, where the idea for this paper was conceived. 
We  thank Hans-Walter Rix for stimulating discussions during the early stages of this work.
 We also thank Adam Ginsburg and Alvaro Hacar for their comments in the later stages of the project, which significantly improved this paper.
This work has made use of data from the European Space Agency (ESA)
mission {\it Gaia} (\url{https://www.cosmos.esa.int/gaia}), processed by
the {\it Gaia} Data Processing and Analysis Consortium (DPAC,
\url{https://www.cosmos.esa.int/web/gaia/dpac/consortium}). Funding
for the DPAC has been provided by national institutions, in particular
the institutions participating in the {\it Gaia} Multilateral Agreement.
This research has made use of the SIMBAD database,
operated at CDS, Strasbourg, France.




\bibliographystyle{mnras}
\bibliography{GAIA_bibliography} 




\appendix

\section{Position-position maps with velocity vectors}

In this Appendix, we present the velocity vector fields for all 18 of the OB associations presented in this work. These maps are directly analogous to Figure \ref{CasOB_exmpl}.

\begin{figure*}
	\begin{minipage}{165mm}
		\begin{center}
			\includegraphics[width=0.99\linewidth]{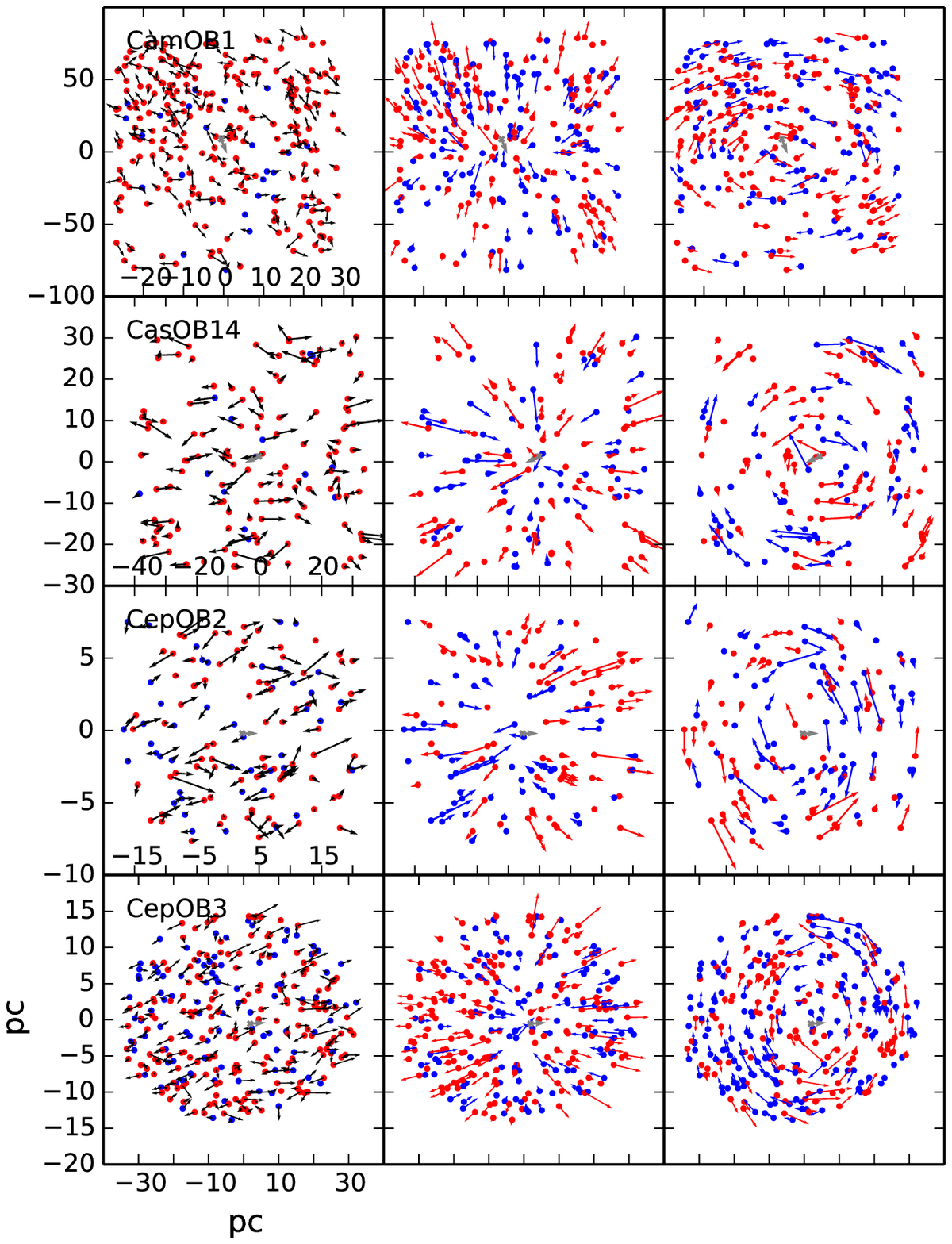}
			\caption{\label{fig:A1} Two-dimensional position-position diagrams for each OB association analysed in this work. The vectors show the observed proper velocities to the mean of all stars (left panel), the radial velocity component with respect to the mean position of all stars (centre panel) and the tangential velocity component with respect to the mean position of all stars (right panel) The mean position and velocity vector of the OB stars relative to the centre of all stars is marked in grey. All distances are in pc.}
		\end{center}
	\end{minipage}
\end{figure*}

\begin{figure*}
	\begin{minipage}{165mm}
		\begin{center}
			\includegraphics[width=0.99\linewidth]{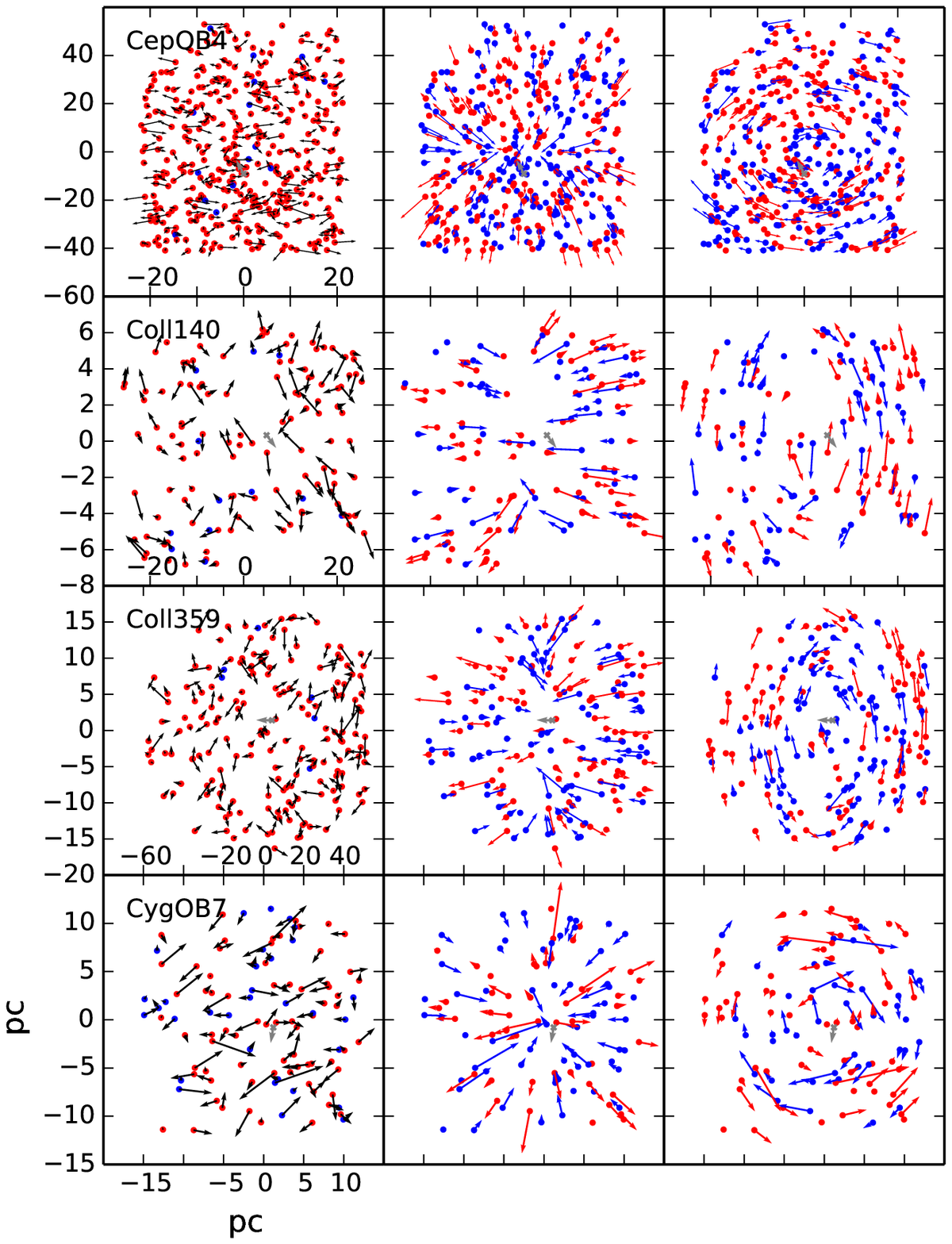}
		\end{center}
			{\bf Figure A1 continued.} Two-dimensional position-position diagrams for each OB association analysed in this work. The vectors show the observed proper velocities to the mean of all stars (left panel), the radial velocity component with respect to the mean position of all stars (centre panel) and the tangential velocity component with respect to the mean position of all stars (right panel) The mean position and velocity vector of the OB stars relative to the centre of all stars is marked in grey. All distances are in pc.
	\end{minipage}
\end{figure*}

\begin{figure*}
	\begin{minipage}{165mm}
		\begin{center}
			\includegraphics[width=0.99\linewidth]{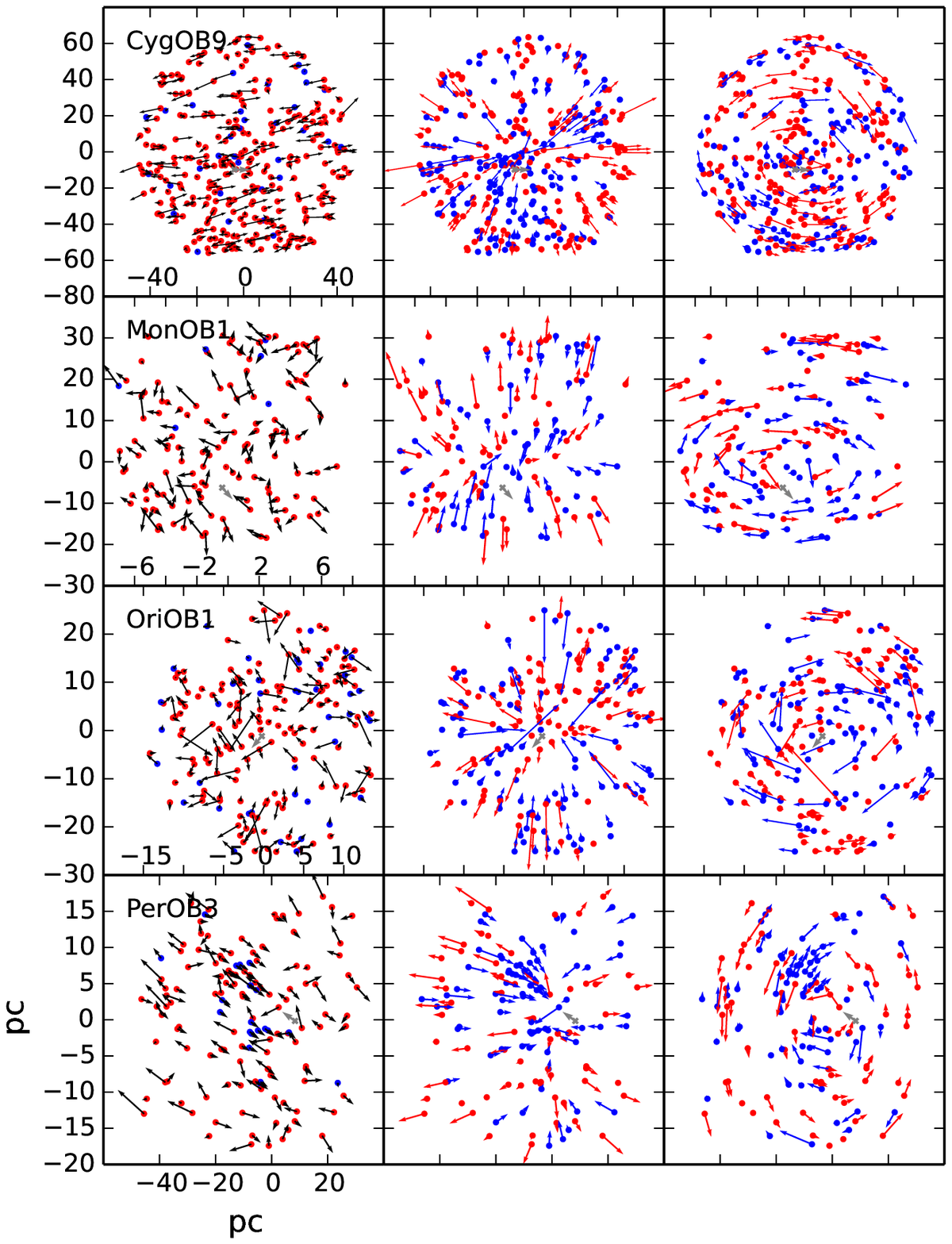}
		\end{center}
			{\bf Figure A1 continued.} Two-dimensional position-position diagrams for each OB association analysed in this work. The vectors show the observed proper velocities to the mean of all stars (left panel), the radial velocity component with respect to the mean position of all stars (centre panel) and the tangential velocity component with respect to the mean position of all stars (right panel) The mean position and velocity vector of the OB stars relative to the centre of all stars is marked in grey. All distances are in pc.
	\end{minipage}
\end{figure*}

\begin{figure*}
	\begin{minipage}{165mm}
		\begin{center}
			\includegraphics[width=0.99\linewidth]{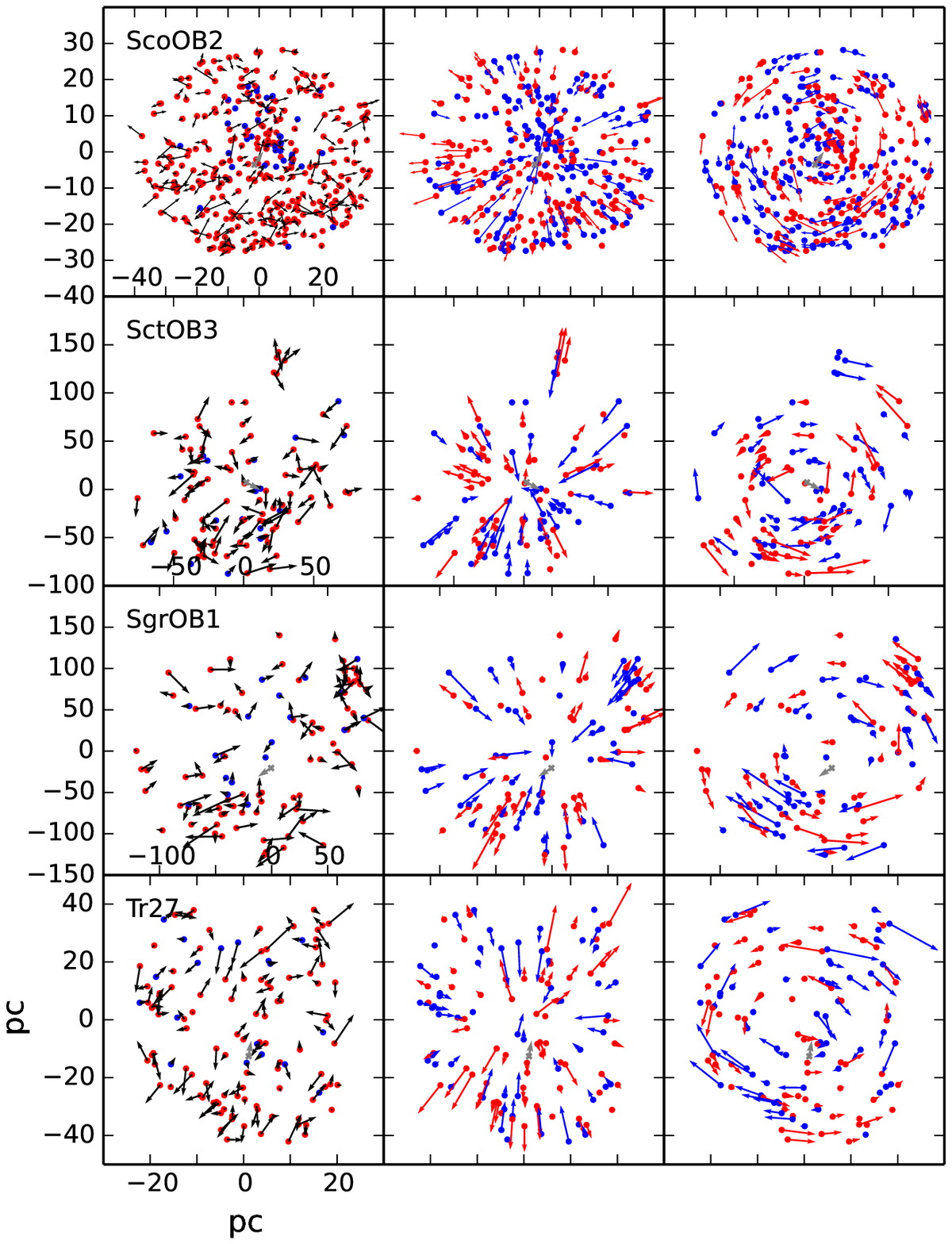}
		\end{center}
		{\bf Figure A1 continued.} Two-dimensional position-position diagrams for each OB association analysed in this work. The vectors show the observed proper velocities to the mean of all stars (left panel), the radial velocity component with respect to the mean position of all stars (centre panel) and the tangential velocity component with respect to the mean position of all stars (right panel) The mean position and velocity vector of the OB stars relative to the centre of all stars is marked in grey. All distances are in pc.
	\end{minipage}
\end{figure*}

\begin{figure*}
	\begin{minipage}{165mm}
		\begin{center}
			\includegraphics[width=0.99\linewidth]{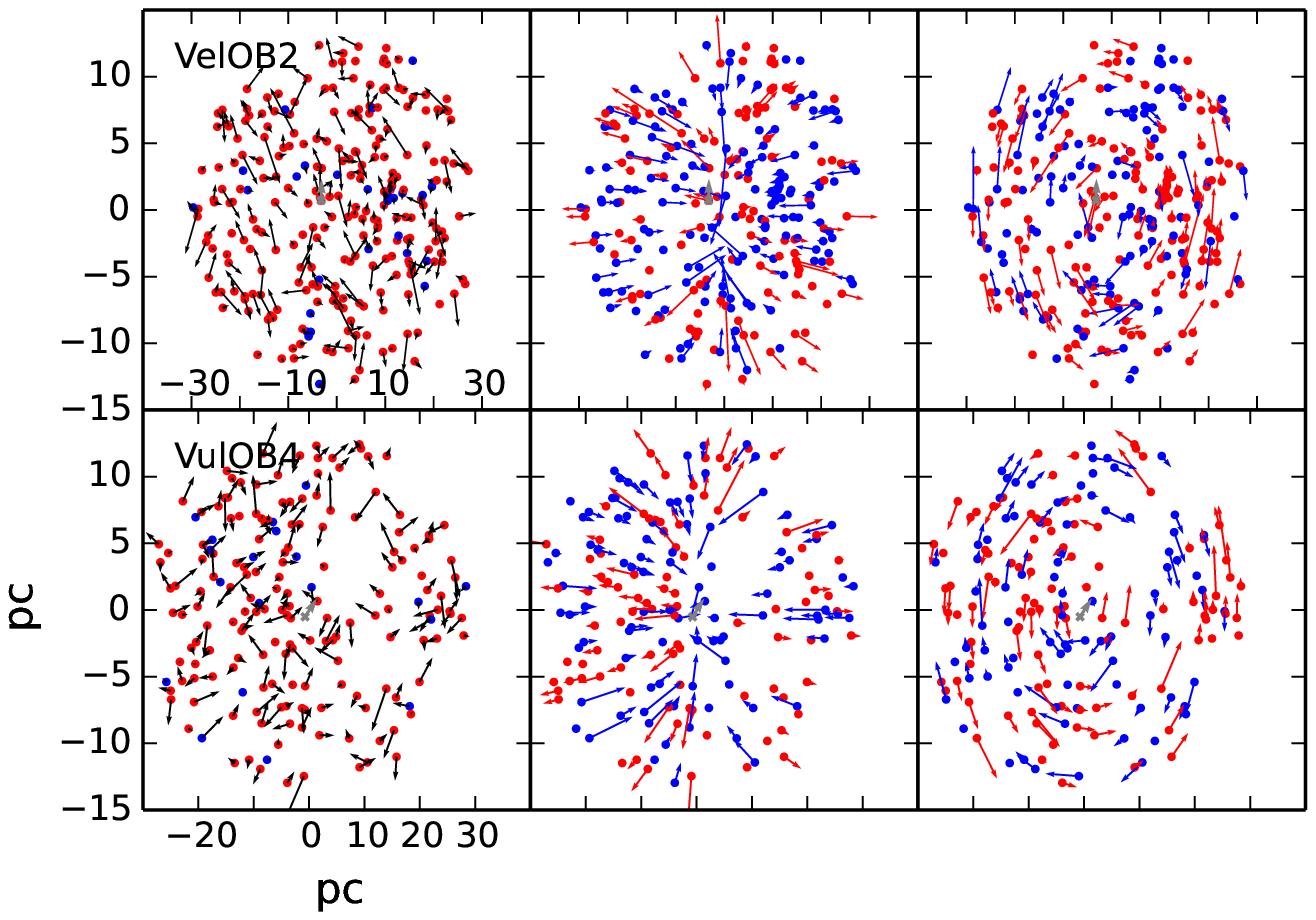}
		\end{center}
		{\bf Figure A1 continued.} Two-dimensional position-position diagrams for each OB association analysed in this work. The vectors show the observed proper velocities to the mean of all stars (left panel), the radial velocity component with respect to the mean position of all stars (centre panel) and the tangential velocity component with respect to the mean position of all stars (right panel) The mean position and velocity vector of the OB stars relative to the centre of all stars is marked in grey. All distances are in pc.
	\end{minipage}
\end{figure*}

\section{OB association members}

The complete list of all OB association members selected following the procedure outlined in Section 2.1 is available online as supplementary material. In Table \ref{CasOB14_allmembers}, we present a small subset (40 stars) of the Cas OB14 association members for reference. For each star we list the association, the Tycho2 identification (or the Hipparcos identification if a Tycho2 ID is not available in the TGAS catalogue), the J2000 RA and Dec, the parallaxes, and the proper motions in RA and Dec. In the final column we mark whether the star has been identified as spectral type O or B from cross-matching the TGAS catalogue data with the Simbad database.

\begin{table*}
	\begin{minipage}{170mm}
		\caption{\label{CasOB14_allmembers} A randomly selected subset of 40 members of the Cas OB14 association, selected following the procedure outlined in Section 2.1. All OB association members selected in this study are listed in the online supporting information. The first column lists the OB association to which the star has been assigned membership. This is followed by the Tycho2 identification as listed in the TGAS catalogue. The third and fourth columns give the RA and Dec coordinates in the J2000 reference frame. The fifth, sixth, and seventh columns list the parallax, proper motions ($\mu$) in RA and Dec as listed in the TGAS catalogue. The final column indicates whether the star is listed as spectral type O or B in the Simbad database.}

\begin{center}
	\begin{tabular}{l c c c c c c c}
		\hline
		OB association & Tycho2 ID & RA (J2000) & Dec (J2000) & parallax (mas) & $\mu_{\text{RA}}$ (mas\,yr$^{-1}$) & $\mu_{\text{Dec}}$ (mas\,yr$^{-1}$) & OB-type star? \\
		\hline
		CasOB14	&	4014-3102-1	&	00 16 37.486	&	+60 53 33.94	&	1.053	$\pm$	0.312	&	\llap{$-$}5.032	$\pm$	0.722	&	\llap{$-$}1.164	$\pm$	0.596	&	\cmark	\\
		CasOB14	&	4015-1652-1	&	00 17 24.908	&	+60 58 25.03	&	0.982	$\pm$	0.293	&	\llap{$-$}5.894	$\pm$	0.787	&	\llap{$-$}1.755	$\pm$	0.668	&		\\
		CasOB14	&	4014-2862-1	&	00 15 47.854	&	+60 40 03.68	&	1.103	$\pm$	0.303	&	\llap{$-$}6.879	$\pm$	1.013	&	\llap{$-$}5.649	$\pm$	0.762	&		\\
		CasOB14	&	4014-1614-1	&	00 16 53.644	&	+61 07 10.30	&	1.132	$\pm$	0.319	&	2.414	$\pm$	0.940	&	\llap{$-$}0.044	$\pm$	0.682	&	\cmark	\\
		CasOB14	&	4014-684-1	&	00 16 47.168	&	+60 25 23.52	&	1.120	$\pm$	0.309	&	1.510	$\pm$	1.148	&	\llap{$-$}0.041	$\pm$	0.627	&		\\
		CasOB14	&	4015-3147-1	&	00 18 10.758	&	+60 32 10.53	&	1.203	$\pm$	0.271	&	\llap{$-$}5.399	$\pm$	0.794	&	\llap{$-$}1.188	$\pm$	0.555	&		\\
		CasOB14	&	4014-2939-1	&	00 13 42.385	&	+60 38 48.52	&	1.265	$\pm$	0.340	&	-5.984	$\pm$	0.869	&	1.405	$\pm$	0.691	&		\\
		CasOB14	&	4014-3112-1	&	00 15 43.344	&	+61 17 15.39	&	1.317	$\pm$	0.352	&	\llap{$-$}1.511	$\pm$	0.904	&	\llap{$-$}1.970	$\pm$	0.640	&		\\
		CasOB14	&	4015-1548-1	&	00 18 53.687	&	+60 43 52.13	&	1.077	$\pm$	0.297	&	\llap{$-$}1.087	$\pm$	0.818	&	\llap{$-$}0.365	$\pm$	0.655	&		\\
		CasOB14	&	4015-2554-1	&	00 18 20.236	&	+60 24 17.02	&	1.138	$\pm$	0.290	&	\llap{$-$}1.139	$\pm$	0.991	&	0.483	$\pm$	0.604	&		\\
		CasOB14	&	4015-1730-1	&	00 17 43.572	&	+61 10 53.17	&	1.264	$\pm$	0.407	&	3.703	$\pm$	1.160	&	0.278	$\pm$	0.930	&		\\
		CasOB14	&	4014-1397-1	&	00 15 59.185	&	+60 02 43.90	&	1.228	$\pm$	0.406	&	5.706	$\pm$	1.117	&	\llap{$-$}0.750	$\pm$	0.828	&		\\
		CasOB14	&	4014-3083-1	&	00 13 25.391	&	+60 47 59.53	&	1.242	$\pm$	0.269	&	\llap{$-$}6.496	$\pm$	0.801	&	1.276	$\pm$	0.565	&		\\
		CasOB14	&	3665-1263-1	&	00 17 08.615	&	+59 59 27.36	&	1.278	$\pm$	0.301	&	4.211	$\pm$	0.782	&	\llap{$-$}0.336	$\pm$	0.641	&		\\
		CasOB14	&	4014-853-1	&	00 15 07.726	&	+61 19 59.36	&	1.119	$\pm$	0.271	&	\llap{$-$}7.133	$\pm$	0.968	&	1.526	$\pm$	0.891	&		\\
		CasOB14	&	4014-2440-1	&	00 13 08.858	&	+60 52 51.75	&	1.260	$\pm$	0.364	&	\llap{$-$}0.982	$\pm$	0.974	&	\llap{$-$}0.282	$\pm$	0.821	&		\\
		CasOB14	&	4014-2720-1	&	00 12 55.921	&	+60 48 04.63	&	1.266	$\pm$	0.327	&	\llap{$-$}2.479	$\pm$	0.934	&	0.584	$\pm$	0.653	&		\\
		CasOB14	&	4014-1882-1	&	00 16 55.695	&	+61 28 53.11	&	1.086	$\pm$	0.331	&	\llap{$-$}4.753	$\pm$	1.334	&	1.260	$\pm$	0.756	&		\\
		CasOB14	&	4015-1098-1	&	00 18 06.903	&	+61 27 51.31	&	1.203	$\pm$	0.313	&	0.385	$\pm$	0.852	&	0.162	$\pm$	0.663	&		\\
		CasOB14	&	4015-2748-1	&	00 18 53.528	&	+60 10 32.76	&	1.127	$\pm$	0.278	&	\llap{$-$}2.038	$\pm$	0.733	&	\llap{$-$}0.901	$\pm$	0.538	&		\\
		CasOB14	&	4015-2775-1	&	00 19 29.015	&	+60 17 13.91	&	1.220	$\pm$	0.371	&	\llap{$-$}1.489	$\pm$	0.944	&	\llap{$-$}2.868	$\pm$	0.832	&		\\
		CasOB14	&	4014-645-1	&	00 12 58.035	&	+60 20 06.27	&	1.214	$\pm$	0.337	&	2.151	$\pm$	0.892	&	\llap{$-$}0.336	$\pm$	0.577	&		\\
		CasOB14	&	4014-303-1	&	00 13 09.0811	&	+60 13 05.16	&	1.216	$\pm$	0.295	&	\llap{$-$}0.448	$\pm$	0.636	&	\llap{$-$}1.657	$\pm$	0.580	&	\cmark	\\
		CasOB14	&	4015-2890-1	&	00 19 51.498	&	+60 25 58.78	&	1.180	$\pm$	0.274	&	\llap{$-$}1.156	$\pm$	1.100	&	\llap{$-$}2.470	$\pm$	0.772	&		\\
		CasOB14	&	4014-1092-1	&	00 14 52.774	&	+61 26 56.03	&	1.038	$\pm$	0.272	&	\llap{$-$}0.273	$\pm$	0.867	&	\llap{$-$}0.676	$\pm$	0.678	&	\cmark	\\
		CasOB14	&	3665-897-1	&	00 16 25.002	&	+59 43 36.08	&	1.238	$\pm$	0.277	&	4.377	$\pm$	0.810	&	\llap{$-$}0.897	$\pm$	0.569	&		\\
		CasOB14	&	4014-339-1	&	00 16 28.753	&	+60 05 08.67	&	0.986	$\pm$	0.292	&	\llap{$-$}1.762	$\pm$	0.777	&	0.581	$\pm$	0.581	&		\\
		CasOB14	&	4015-2837-1	&	00 20 13.930	&	+61 00 12.69	&	1.259	$\pm$	0.360	&	\llap{$-$}0.509	$\pm$	0.912	&	0.320	$\pm$	0.729	&		\\
		CasOB14	&	3664-20-1	&	00 12 33.634	&	+59 57 31.92	&	1.300	$\pm$	0.407	&	\llap{$-$}5.957	$\pm$	1.009	&	0.842	$\pm$	0.794	&		\\
		CasOB14	&	4015-1824-1	&	00 20 21.159	&	+60 25 47.84	&	1.321	$\pm$	0.290	&	\llap{$-$}3.278	$\pm$	0.839	&	\llap{$-$}0.724	$\pm$	0.531	&	\cmark	\\
		CasOB14	&	4014-473-1	&	00 17 01.460	&	+61 43 36.73	&	1.046	$\pm$	0.335	&	\llap{$-$}2.025	$\pm$	0.849	&	\llap{$-$}0.342	$\pm$	0.675	&		\\
		CasOB14	&	4014-1727-1	&	00 15 02.242	&	+60 11 05.70	&	0.977	$\pm$	0.315	&	\llap{$-$}1.642	$\pm$	0.874	&	0.989	$\pm$	0.654	&		\\
		CasOB14	&	4014-711-1	&	00 13 44.468	&	+60 12 17.27	&	1.076	$\pm$	0.329	&	\llap{$-$}4.475	$\pm$	0.818	&	\llap{$-$}1.405	$\pm$	0.580	&		\\
		CasOB14	&	4014-701-1	&	00 14 43.922	&	+61 46 14.97	&	1.225	$\pm$	0.289	&	\llap{$-$}5.762	$\pm$	0.685	&	\llap{$-$}3.222	$\pm$	0.527	&		\\
		CasOB14	&	4014-1433-1	&	00 10 52.554	&	+60 26 32.03	&	1.321	$\pm$	0.379	&	\llap{$-$}0.380	$\pm$	0.721	&	\llap{$-$}0.220	$\pm$	0.605	&	\cmark	\\
		CasOB14	&	4015-2732-1	&	00 20 47.407	&	+60 29 46.06	&	0.979	$\pm$	0.265	&	0.128	$\pm$	0.722	&	\llap{$-$}2.207	$\pm$	0.532	&		\\
		CasOB14	&	3665-43-1	&	00 16 02.327	&	+59 55 05.71	&	0.998	$\pm$	0.310	&	\llap{$-$}2.675	$\pm$	0.823	&	\llap{$-$}0.602	$\pm$	0.967	&	\cmark	\\
		CasOB14	&	4018-1953-1	&	00 16 13.288	&	+61 53 33.19	&	1.099	$\pm$	0.269	&	0.847	$\pm$	0.579	&	\llap{$-$}5.900	$\pm$	0.506	&		\\
		CasOB14	&	4015-3109-1	&	00 20 24.791	&	+60 13 34.61	&	1.000	$\pm$	0.292	&	\llap{$-$}2.983	$\pm$	0.787	&	0.819	$\pm$	0.636	&		\\
		CasOB14	&	3665-81-1	&	00 19 40.617	&	+59 44 16.38	&	1.266	$\pm$	0.254	&	1.755	$\pm$	0.643	&	\llap{$-$}3.159	$\pm$	0.504	&		\\
		\hline
	\end{tabular}
\end{center}
\end{minipage}
\end{table*}

\bsp	
\label{lastpage}
\end{document}